# Dissecting the NVIDIA Volta GPU Architecture via Microbenchmarking

*Technical Report*

*First Edition*
*April 18th, 2018*

Zhe Jia
Marco Maggioni
Benjamin Staiger
Daniele P. Scarpazza

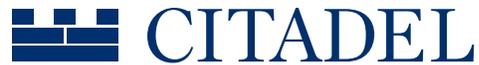

The authors make this edition, and will make future editions, of this document available on the arXiv.org e-print service owned and operated by Cornell University.

| Edition | Date |
| --- | --- |
| First | April 18th, 2018 |

The lead author, Zhe Jia, presented a limited preview of the material contained in the first edition of this report at the 2018 GPU Technology Conference, March 26-29, 2018, San Jose, California.

The authors welcome all feedback. The corresponding author is Daniele P. Scarpazza. Send feedback preferably to his e-mail address: daniele.scarpazza@citadel.com.

The authors are grateful to NVIDIA for their support and cooperation.

The authors thank their colleague Jeffrey T. Smith for his technical insights, and their colleague Julia Kosygina for her editorial support.

# Contents

# Summary

Every year, novel NVIDIA® GPU designs are introduced [1, 2, 3, 4, 5, 6]. This rapid architectural and technological progression, coupled with a reluctance by manufacturers to disclose low-level details, makes it difficult for even the most proficient GPU software designers to remain up-to-date with the technological advances at a microarchitectural level.

To address this dearth of public, microarchitectural-level information on the novel NVIDIA GPUs, independent researchers have resorted to microbenchmarks-based dissection and discovery. This has led to a prolific line of publications that shed light on instruction encoding [7, 8, 9], and memory hierarchy's geometry and features [10] at each level. Namely, research that describes the performance and behavior of the Kepler [11], Maxwell [12] and Pascal architectures.

In this technical report, we continue this line of research by presenting the microarchitectural details of the NVIDIA Volta architecture, discovered through microbenchmarks and instruction set disassembly.

Additionally, we compare quantitatively our Volta findings against its predecessors, Kepler, Maxwell and Pascal.

# Chapter 1

# Why these details matter

In this chapter, we show that deep, architectural understanding is necessary to optimize software to its peak performance, and that the amount of performance left on the table without this understanding can be substantial.

Published work, together with our experiments, consistently show that bare-metal peak performance is inaccessible to software written in plain CUDA and without a deeper understanding of the hardware compared to what is currently publicly available. There is consensus that the high degree of optimization found in NVIDIA's libraries such as cuBlas and cuDNN is inaccessible to authors of CUDA code, even when they use inline PTX assembly. We believe that manufacturers write optimized libraries as low-level SASS code, possibly with assemblers which they do not make available to the public.

Multiple authors [9, 13, 14, 15, 16, 17] have shown areas for improved efficiency in the machine code emitted by the NVIDIA CUDA Compiler (NVCC) for architectures preceding Volta. Authors have also shown that knowledge of instruction encoding and microarchitectural behavior is necessary to achieve this full potential.

We extend this line of research to Volta: we show that opportunities for improving performance still exist in code emitted by the NVCC compiler for the Volta architecture, and we demonstrate how to exploit them with the help of a minimal example.

However, our approach has limitations:

- the optimizations we describe require substantial human effort. This effort may not, in general, be worth the gains it produces, except for tight computational kernels that vastly dominate an application's execution time;

- our optimizations are specific to the Volta architecture.  Their perfor-
  mance gains will not port to future GPU architectures.  Our discovery
  work needs to be repeated anew on each future architecture;

- developers that use the CUDA libraries and the NVCC compiler with-
  out our optimizations already benefit from an excellent combination of
  portability and efficiency.  They will, in general, benefit from perfor-
  mance gains every time a new GPU generation is introduced, at little
  or no extra development effort.

The simplest, meaningful example we could construct to illustrate our
claims is the following kernel, representative of the innermost core of a fixed-
size, matrix-matrix multiplication:

```
__shared__ float        shr_A[8*512], shr_B[8*512];
float                   reg_A[8], reg_B[8], reg_C[64];
// ...

for (int k=0; k<512; k++) {
    for (int i = 0; i<8; i++) {
        reg_A[i]=shr_A[i+k*8];
        reg_B[i]=shr_B[i+k*8];
    }

    for (int i = 0; i<8; i++)
        for (int j = 0; j<8; j++)
            reg_C[i*8+j] += reg_A[i]*reg_B[j];
}
// ...
```

This code multiplies an 8-value row slice from input matrix A against an 8-
value column slice from input matrix B. It first copies the two slices to register-
mapped arrays `reg_A` and `reg_B`. the multiplication accumulates its results into
register-mapped matrix tile `reg_C`, of size 8×8.

This example is compiled into machine code, using NVCC 9.0. The result-
ing machine code exhibits a certain number of *register bank conflicts* that could
be removed; there are also opportunities to better use the *register reuse cache* to
eliminate these conflicts. We list this machine encode, in its entirety, in the left
column of Table 1.1.

We must now explain the details of register bank conflicts, why they are
harmful, and how to mitigate them. On Volta, registers are divided into two
64-bit wide banks. One Volta instruction can only access 64 bits of each bank
per clock cycle. Thus an instruction like FFMA (single precision floating-point
fused multiply-add operation) can read at most two values from each bank
per clock. Any FFMA instruction that accesses the same bank with all its 3
source registers is said to have a *bank conflict*. This means the architecture
cannot satisfy the accesses simultaneously; instead, they must be serialized
which takes longer and is inefficient.

As an example, consider the following four instructions from the machine
code listed. They all suffer from conflicts, described in the comments:

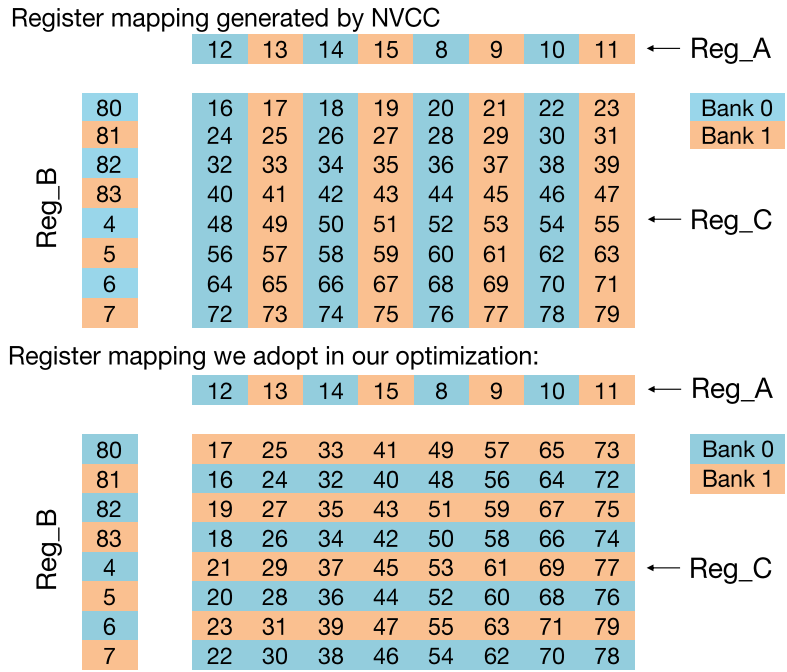

Figure 1.1: Comparison between the register mapping generated by NVCC 9.0 for the motivational example, and the optimized version we adopt.

```
...
FFMA R16, R12, R80, R16;        # R16, R12 and R80 are in bank 0
...
FFMA R25, R13, R81, R25;        # R13, R81 and R25 are in bank 1
...
FFMA R31, R11, R81, R31;        # R11, R81 and R31 are in bank 1
...
FFMA R43, R15, R83, R43;        # R43, R83 and R43 are in bank 1
...
```

We rewrite the code at the binary level so that it uses a better register mapping, suffers no conflicts and leverages register reuse caches. The effect of our changes are visible in Figure 1, where we color-code the registers by bank they belong to. The top row (`Reg_A`) and the leftmost column (`Reg_B`) show which registers correspond to the input slices from matrix A and B. Each cell in matrix (`Reg_C`) shows the target register where the product of the row and column operand is accumulated. The very first cell indicates that the product between R12 and R80 is accumulated to R16. This instruction suffers from poor register mapping because all three registers belong to Bank 0.

The complete machine code, before and after our optimization, is in Table 1.1. When using 128 threads, our changes improve performance from

132.05 to 152.43 GFlops/s per SMX (+15.4%).

The bottom half of Figure 1 shows our improved mapping. On previous architectures with 32-bit wide banks (Maxwell [9] and Pascal), one can only access one value from each memory bank per clock cycle. This means register mapping cannot eliminate conflicts between row and column input registers that belong to the same bank. One can only mitigate those conflicts via register reuse caches. However, on Volta, even when row and column input registers belong to the same bank, one can avoid the conflict by using a third input register from the other bank.

The task of patching machine code to eliminate conflicts, as we just illustrated, would be impossible if we didn't first discover the instruction encoding format and study the behavior of bank conflicts. Both these topics are primary contributions of this work.

Table 1.1: Comparison between the machine code generated by NVCC when compiling the motivational example and machine code we improved. Our code features register mapping that avoids bank conflicts and reuses register caches as much as possible.

| Original machine code generated by NVCC 9.0 | Our improved machine code |
| --- | --- |
| FFMA R16, R12, R80, R16; | FFMA R17, R12.reuse, R80.reuse, R17; |
| FFMA R17, R80.reuse, R13, R17; | FFMA R16, R12, R81.reuse, R16; |
| FFMA R18, R80.reuse, R14, R18; | FFMA R25, R13.reuse, R80.reuse, R25; |
| FFMA R19, R80, R15, R19; | FFMA R24, R13, R81.reuse, R24; |
| FFMA R20, R80.reuse, R8, R20; | FFMA R33, R14.reuse, R80.reuse, R33; |
| FFMA R21, R80.reuse, R9, R21; | FFMA R32, R14, R81.reuse, R32; |
| FFMA R22, R80.reuse, R10, R22; | FFMA R41, R15.reuse, R80.reuse, R41; |
| FFMA R23, R80, R11, R23; | FFMA R40, R15, R81.reuse, R40; |
| FFMA R24, R12, R81.reuse, R24; | FFMA R49, R8.reuse, R80.reuse, R49; |
| FFMA R25, R13, R81, R25; | FFMA R48, R8, R81.reuse, R48; |
| FFMA R26, R14, R81.reuse, R26; | FFMA R57, R9.reuse, R80.reuse, R57; |
| FFMA R27, R15, R81.reuse, R27; | FFMA R56, R9, R81.reuse, R56; |
| FFMA R28, R8, R81.reuse, R28; | FFMA R65, R10.reuse, R80.reuse, R65; |
| FFMA R29, R9, R81.reuse, R29; | FFMA R64, R10.reuse, R81.reuse, R64; |
| FFMA R30, R10, R81.reuse, R30; | FFMA R73, R11.reuse, R80, R73; |
| FFMA R31, R11, R81, R31; | FFMA R72, R11.reuse, R81, R72; |
| FFMA R32, R12, R82.reuse, R32; | FFMA R75, R11.reuse, R82.reuse, R75; |
| FFMA R33, R13, R82.reuse, R33; | FFMA R74, R11, R83.reuse, R74; |
| FFMA R34, R14, R82.reuse, R34; | FFMA R67, R10.reuse, R82.reuse, R67; |
| FFMA R35, R15, R82.reuse, R35; | FFMA R66, R10, R83.reuse, R66; |
| FFMA R36, R8, R82.reuse, R36; | FFMA R59, R9.reuse, R82.reuse, R59; |
| FFMA R37, R9, R82, R37; | FFMA R58, R9, R83.reuse, R58; |
| FFMA R38, R10, R82.reuse, R38; | FFMA R51, R8.reuse, R82.reuse, R51; |
| FFMA R39, R11, R82, R39; | FFMA R50, R8, R83.reuse, R50; |
| FFMA R40, R12, R83.reuse, R40; | FFMA R43, R15.reuse, R82.reuse, R43; |
| FFMA R41, R13, R83.reuse, R41; | FFMA R42, R15, R83.reuse, R42; |
| FFMA R42, R14, R83.reuse, R42; | FFMA R35, R14.reuse, R82.reuse, R35; |
| FFMA R43, R15, R83, R43; | FFMA R34, R14, R83.reuse, R34; |
| FFMA R44, R8, R83.reuse, R44; | FFMA R27, R13.reuse, R82.reuse, R27; |
| FFMA R45, R9, R83.reuse, R45; | FFMA R26, R13.reuse, R83.reuse, R26; |
| FFMA R46, R10, R83.reuse, R46; | FFMA R19, R12.reuse, R82, R19; |
| FFMA R47, R11, R83, R47; | FFMA R18, R12.reuse, R83, R18; |
| FFMA R48, R12, R4.reuse, R48; | FFMA R21, R12.reuse, R4.reuse, R21; |
| FFMA R49, R13, R4, R49; | FFMA R20, R12, R5.reuse, R20; |
| FFMA R50, R14, R4.reuse, R50; | FFMA R29, R13.reuse, R4.reuse, R29; |
| FFMA R51, R15, R4.reuse, R51; | FFMA R28, R13, R5.reuse, R28; |
| FFMA R52, R8, R4.reuse, R52; | FFMA R37, R14.reuse, R4.reuse, R37; |
| FFMA R53, R9, R4.reuse, R53; | FFMA R36, R14, R5.reuse, R36; |
| FFMA R54, R10, R4.reuse, R54; | FFMA R45, R15.reuse, R4.reuse, R45; |
| FFMA R55, R11, R4, R55; | FFMA R44, R15, R5.reuse, R44; |
| FFMA R56, R12, R5.reuse, R56; | FFMA R53, R8.reuse, R4.reuse, R53; |
| FFMA R57, R13, R5.reuse, R57; | FFMA R52, R8, R5.reuse, R52; |
| FFMA R58, R14, R5.reuse, R58; | FFMA R61, R9.reuse, R4.reuse, R61; |
| FFMA R59, R15, R5.reuse, R59; | FFMA R60, R9, R5.reuse, R60; |
| FFMA R60, R8, R5.reuse, R60; | FFMA R69, R10.reuse, R4.reuse, R69; |
| FFMA R61, R9, R5, R61; | FFMA R68, R10.reuse, R5.reuse, R68; |
| FFMA R62, R10, R5.reuse, R62; | FFMA R77, R11.reuse, R4, R77; |
| FFMA R63, R11, R5, R63; | FFMA R76, R11.reuse, R5, R76; |
| FFMA R64, R12, R6.reuse, R64; | FFMA R79, R11.reuse, R6.reuse, R79; |
| FFMA R65, R13, R6.reuse, R65; | FFMA R78, R11, R7.reuse, R78; |
| FFMA R66, R14, R6.reuse, R66; | FFMA R71, R10.reuse, R6.reuse, R71; |
| FFMA R67, R15, R6, R67; | FFMA R70, R10, R7.reuse, R70; |
| FFMA R68, R8, R6.reuse, R68; | FFMA R63, R9.reuse, R6.reuse, R63; |
| FFMA R69, R9, R6.reuse, R69; | FFMA R62, R9, R7.reuse, R62; |
| FFMA R70, R10, R6.reuse, R70; | FFMA R55, R8.reuse, R6.reuse, R55; |
| FFMA R71, R11, R6, R71; | FFMA R54, R8, R7.reuse, R54; |
| FFMA R72, R12, R7.reuse, R72; | FFMA R47, R15.reuse, R6.reuse, R47; |
| FFMA R73, R13, R7, R73; | FFMA R46, R15, R7.reuse, R46; |
| FFMA R74, R14, R7.reuse, R74; | FFMA R39, R14.reuse, R6.reuse, R39; |
| FFMA R75, R15, R7.reuse, R75; | FFMA R38, R14, R7.reuse, R38; |
| FFMA R76, R8, R7.reuse, R76; | FFMA R31, R13.reuse, R6.reuse, R31; |
| FFMA R77, R9, R7.reuse, R77; | FFMA R30, R13.reuse, R7.reuse, R30; |
| FFMA R78, R10, R7.reuse, R78; | FFMA R23, R12.reuse, R6, R23; |
| FFMA R79, R11, R7, R79; | FFMA R22, R12.reuse, R7, R22; |

# Chapter 2

# Instructions

Volta adopts a different instruction encoding from that used on the Pascal and Maxwell architectures.

Volta uses one 128-bit word to encode each instruction together with its corresponding control information. This is a substantial departure from previous architectures that use a 64-bit word to encode each instruction, and a separate 64-bit word to encode control information associated to multiple instructions. The following example illustrates a Volta instruction, together with its control information, as disassembled by `nvdisasm` [18]:

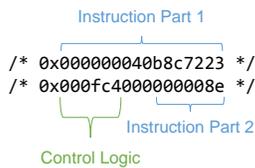

`nvdisasm` shows the 128-bit word as two 64-bit words. The first word only encodes instruction information. The second encodes both instruction and control information.

To the best of the knowledge that we were able to derive from exhaustive instruction disassembly, the 128 bits in a word are used as follows:

- at least 91 bits are used to encode the instruction;

- at least 23 bits are used to encode control information;

- the remaining 14 bits appeared to be unused in our experiments.

## 2.1   Control information

Kepler introduced control words to encode instruction scheduling decisions taken by the compiler [11]. Control information prevents data hazards and allows for simpler on-chip control logic, leading to higher compute density per area of silicon and better energy efficiency.

On Volta, a single 128-bit word contains one instruction together with the control information associated to that instruction.

Pre-Volta architectures interleave and bundle control words with instructions. In each bundle, the first word of every bundle is a control word, and the remaining words (3 on Pascal and Maxwell, 7 on Kepler) are instructions. Each control word affects how the architecture schedules the instructions within the bundle. The following listing shows an example of a bundle of Pascal instructions, as disassembled by `nvdisasm`. The bundle contains four 64-bit words. The first word, which has a hexadecimal dump but no corresponding disassembled instruction, is a control word. The remaining three words are instructions.

```
/*0288*/        @P5 LDG.E.CI R66, [R86+0x100];        /* 0x000f8800fe2007f1 */
/*0290*/        @!P5 MOV R66, RZ;                      /* 0xeed4a00010055642 */
/*0290*/        @!P5 MOV R66, RZ;                      /* 0x5c9807800ffd0042 */
/*0298*/        @P6 LDG.E.CI R67, [R86+0x180];        /* 0xeed4a00018065643 */
```

Control information is encoded as follows on the different GPU generations:

- on Kepler, each control word contains 6 zeroes as its most significant bits, 2 zeroes as its least significant bits, and 7 sections of 8 bits each;

- on Pascal and Maxwell, each control word contains one zero as its most significant bit, and 3 sections of 21 bits each;

- on Volta, each control section contains 2 zeroes as its most significant bits, and 1 section of 21 bits. In each 128-bit word, control information is preceded and followed by the instruction encoding bits.

Sections containing control information are organized in a similar way on Volta, Pascal and Maxwell. Each section contains 6 fields, organized as follows:

| Width (bits) | 4 | 6 | 3 | 3 | 1 | 4 |
|---|---|---|---|---|---|---|
| Meaning | Reuse flags | Wait barrier mask | Read barrier index | Write barrier index | Yield flag | Stall cycles |

Fields have the following meaning:

**Reuse flags.** Volta, Pascal and Maxwell have 4 register reuse caches and 4 source operand slots. Each of the 4 reuse flag bits corresponds to one of the 8-byte slots. When a flag is set, the value of the register in the corresponding slot will be stored in the reuse cache for future instructions to consume. Reuse mitigates register bank conflicts. The least significant bit in reuse flags controls the cache for the first source operand slot. The most significant bit is for the fourth source operand slot.

**Wait barrier mask; Read/Write barrier index.** While most instructions have fixed latency and can be statically scheduled by the assembler, instructions involving memory and shared resources typically have variable latency. Volta, Pascal and Maxwell use *dependency barriers* to track the completion of variable-latency instructions and resolve data hazards. When a variable-latency instruction writes to a register, the assembler associates it to one of the 6 available barriers by setting the corresponding *write barrier number* field. When a later instruction consumes that register, the assembler marks the instruction as waiting on that barrier by setting the bit corresponding to that barrier in the *wait barrier mask*. The hardware will stall the later instruction until the results of the earlier one are available. An instruction may wait on multiple barriers, which explains why the *wait barrier mask* is a bitmask, not an index.

**Read dependency barriers.** Read dependency barriers serve to protect against write-after-read hazards. Unbuffered instructions that write the contents of registers to memory need the registers to remain unchanged during the operation. To guarantee that, the assembler associates them to a barrier by populating the corresponding *read barrier number* field. Later instructions writing to the same register will wait on that barrier.

**Stall cycles.** This 4-bit field indicates how long the scheduler should wait before issuing the next instruction, ranging from 0 to 15 cycles. On Pascal and Maxwell, if the combination of this field and the yield flag contain a special combination of bits, the two dispatchers in a processing block can dispatch two consecutive instructions of a warp at the same time (dual issue). On Volta there is only one dispatcher in a processing block, and we do not observe dual issue in the generated code.

**Yield flag.** As its predecessors, the Volta architecture uses a one-bit yield flag to balance the workload assigned to a processing block. When this bit is set, the scheduler prefers to issue the next instruction from the current warp. When the bit is cleared, the scheduler prefers to switch to another warp, making all register reuse flags for the next instruction ineffective. This costs one extra cycle to switch to another warp.

## 2.2   Scheduler

The Volta streaming multiprocessor (SM) is partitioned into four processing blocks [6]. Instructions from the same warp are allocated to a specific processing block, and can only access the processing units within that block.

The mapping between the scheduler and the warp index on Volta is as simple as $scheduler\_id = warp\_id\%4$. To prove it, we use a benchmark that executes a sequence of FFMA instructions in two warps on a single SM simultaneously. The benchmark invokes 8 warps every time, of which 6 are idle and the remaining 2 (warp A and warp B) issue FFMA instructions. We change the combination of warps which issue FFMA instructions and measure the achieved throughput in GFLOPS. As shown in Table 2.1, when all FFMA instructions are processed on a single processing block, the measured performance is lower.

This also indicates that at least 128 threads are required to fully use the processing units on Volta.

Table 2.1: Single-precision arithmetic throughput measured while varying the indices of warps used to process FFMA sequences. The benchmark runs on different processing blocks on one Volta SM. All values are in GFLOPS.

|              |   | Warp A Index | | | |
|--------------|---|-------|-------|-------|-------|
|              |   | 0     | 1     | 2     | 3     |
| Warp B Index | 4 | 42.27 | 66.05 | 66.04 | 65.29 |
|              | 5 | 66.05 | 41.98 | 66.04 | 66.04 |
|              | 6 | 66.02 | 66.04 | 42.06 | 66.04 |
|              | 7 | 66.04 | 66.04 | 66.02 | 42.08 |

## 2.3   Instruction encoding

Volta uses more bits to encode its instructions than previous architectures.

Unlike previous architectures (Pascal, Maxwell and Kepler), which organize the opcode in the most significant bits of the instruction, the Volta architecture places the opcode in the least significant bits of the first 64-bit part of the code bundle. We report an opcode reference for Pascal and Volta in the Appendix.

Volta opcodes vary in length from 10 to 13 bits.

As in previous architectures, operands on Volta can be registers (general purpose, special or predication), memory addresses (constant, shared or global), or an immediate value. Predication is regulated by 4 bits: the first bit is a negation flag, and the remaining 3 bits encode a predicate register index.

# Chapter 3

# Memory hierarchy

NVIDIA GPUs increase in complexity at each newer generation. Gaining a deep understanding of GPU memory hierarchy as they evolve is necessary to write efficient code. It is especially important to know the size of each cache memory level, whether that memory is co-located with another cache that might evict its contents, and whether each cache memory is private to a streaming multiprocessor or shared among all.

In this chapter, we describe the structure of Volta's memory hierarchy in detail, depicted in Figure 3.1. Specifically, we reveal:

- the geometry, properties and performance of all cache levels and Translation Look-aside Buffers TLBs;

- register file banks and their conflicts;

- the performance of shared and global memory under load.

Table 3.1 summarize our findings, comparing Volta against Pascal, Maxwell and Kepler.

As published by NVIDIA [6], the V100 GPU employs HBM2 memory, which offers a bandwidth of 900 GB/s (at 877 MHz), in conjunction with a L2 cache of 6,144 kibibyte. Data loaded from global memory is implicitly cached in L1 and L2.

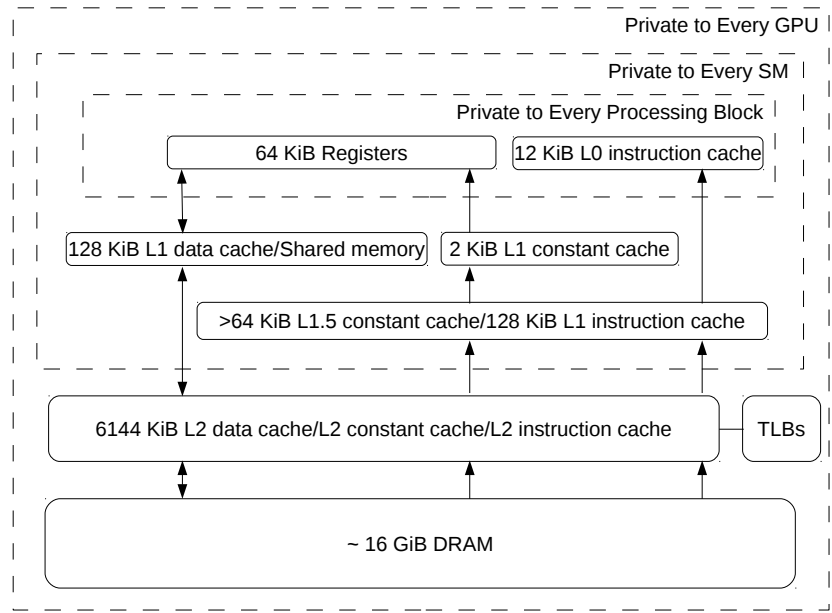

Figure 3.1: Memory hierarchy of the Volta V100 GPU (GV100).

Table 3.1: Geometry, properties and latency of the memory hierarchy on the Volta, Pascal, Maxwell and Kepler architectures. All data in this table are measured on PCI-E cards.

| | | Volta V100 GV100 | Pascal P100 GP100 | Pascal P4 GP104 | Maxwell M60 GM204 | Kepler K80 GK210 |
|---|---|---|---|---|---|---|
| Registers | Number of banks | 2 | 4 | 4 | 4 | 4 |
| | bank width | 64 bit | 32 bit | 32 bit | 32 bit | 32 bit |
| L1 data | Size | 32...128 KiB | 24 KiB | 24 KiB | 24 KiB | 16...48 KiB |
| | Line size | 32 B | 32 B | 32 B | 32 B | 128 B |
| | Hit latency | 28 | 82 | 82 | 82 | 35 |
| | Number of sets | 4 | 4 | 4 | 4 | 32 or 64* |
| | Load granularity | 32 B | 32 B | 32 B | 32 B | 128 B |
| | Update granularity | 128 B | 128 B | 128 B | 128 B | 128 B |
| | Update policy | non-LRU | LRU | LRU | LRU | non-LRU |
| | Physical address indexed | no | no | no | no | no |
| L2 data | Size | 6,144 KiB | 4,096 KiB | 2,048 KiB | 2,048 KiB | 1,536 KiB |
| | Line size | 64 B | 32 B | 32 B | 32 B | 32 B |
| | Hit latency | ~193 | ~234 | ~216 | ~207 | ~200 |
| | Populated by cudaMemcpy | yes | yes | yes | yes | yes |
| | Physical address indexed | yes | yes | yes | yes | yes |
| L1 constant | Broadcast latency | ~27 | ~24 | ~25 | ~25 | ~30 |
| | Cache size | 2 KiB | 2 KiB | 2 KiB | 2 KiB | 2 KiB |
| | Line size | 64 B | 64 B | 64 B | 64 B | 64 B |
| | Number of sets | 8 | 8 | 8 | 8 | 8 |
| | Associativity | 4 | 4 | 4 | 4 | 4 |
| L1.5 constant | Broadcast latency | ~89 | ~96 | ~87 | ~81 | ~92 |
| | Cache size | >=64 KiB | >=64 KiB | 32 KiB | 32 KiB | 32 KiB |
| | Line size | 256 B | 256 B | 256 B | 256 B | 256 B |
| L2 constant | Broadcast latency | ~245 | ~236 | ~225 | ~221 | ~220 |
| L0 instruction | Cache size | ~12 KiB | - | - | - | - |
| L1 instruction | Cache size | 128 KiB | 8 KiB | 8 KiB | 8 KiB | 8 KiB |
| L1.5 instruction | Cache size | - | 128 KiB | 32 KiB | 32 KiB | 32 KiB |
| | SMX private or shared | - | private | private | private | private |
| L2 instruction | Cache size | 6,144 KiB | 4,096 KiB | 2,048 KiB | 2,048 KiB | 1,536 KiB |
| L1 TLB | Coverage | 32 MiB | ~32 MiB | ~32 MiB | ~2 MiB | ~2 MiB |
| | Page entry | 2 MiB | 2 MiB | 2 MiB | 128 KiB | 128 KiB |
| L2 TLB | Coverage | ~8,192 MiB | ~2,048 MiB | ~2,048 MiB | ~128 MiB | ~128 MiB |
| | Page entry | 32 MiB | 32 MiB | 32 MiB | 2 MiB | 2 MiB |
| L3 TLB | Coverage | - | - | - | ~2,048 MiB | ~2,048 MiB |
| | Page entry | - | - | - | 2 MiB | 2 MiB |
| Specifications | Processors per chip (P) | 80 | 56 | 20 | 16 | 13 |
| | Max graphics clock (fg) | 1,380 MHz | 1,328 MHz | 1,531 MHz | 1,177 MHz | 875 MHz |
| Shared memory | Size per SMX | up to 96 KiB | 64 KiB | 64 KiB | 96 KiB | 48 KiB |
| | Size per chip | up to 7,689 KiB | 3,584 KiB | 1,280 KiB | 1,536 KiB | 624 KiB |
| | Banks per processor (Bs) | 32 | 32 | 32 | 32 | 32 |
| | Bank width (ws) | 4 B | 4 B | 4 B | 4 B | 8 B |
| | No-conflict latency | 19 | 24 | 23 | 23 | 26 |
| | Theoretical bandwidth | 13,800 GiB/s | 9,519 GiB/s | 3,919 GiB/s | 2,410 GiB/s | 2,912 GiB/s |
| | Measured bandwidth | 12,080 GiB/s | 7,763 GiB/s | 3,555 GiB/s | 2,122 GiB/s | 2,540 GiB/s |
| Global memory | Memory bus | HBM2 | HBM2 | GDDR5 | GDDR5 | GDDR5 |
| | Size | 16,152 MiB | 16,276 MiB | 8,115 MiB | 8,155 MiB | 12,237 MiB |
| | Max clock rate (fm) | 877 MHz | 715 MHz | 3,003 MHz | 2,505 MHz | 2,505 MHz |
| | Theoretical bandwidth | 900 GiB/s | 732 GiB/s | 192 GiB/s | 160 GiB/s | 240 GiB/s |
| | Measured bandwidth | 750 GiB/s | 510 GiB/s | 162 GiB/s | 127 GiB/s | 191 GiB/s |
| | Measured/Theoretical Ratio | 83.3% | 69.6% | 84.4% | 79.3% | 77.5% |

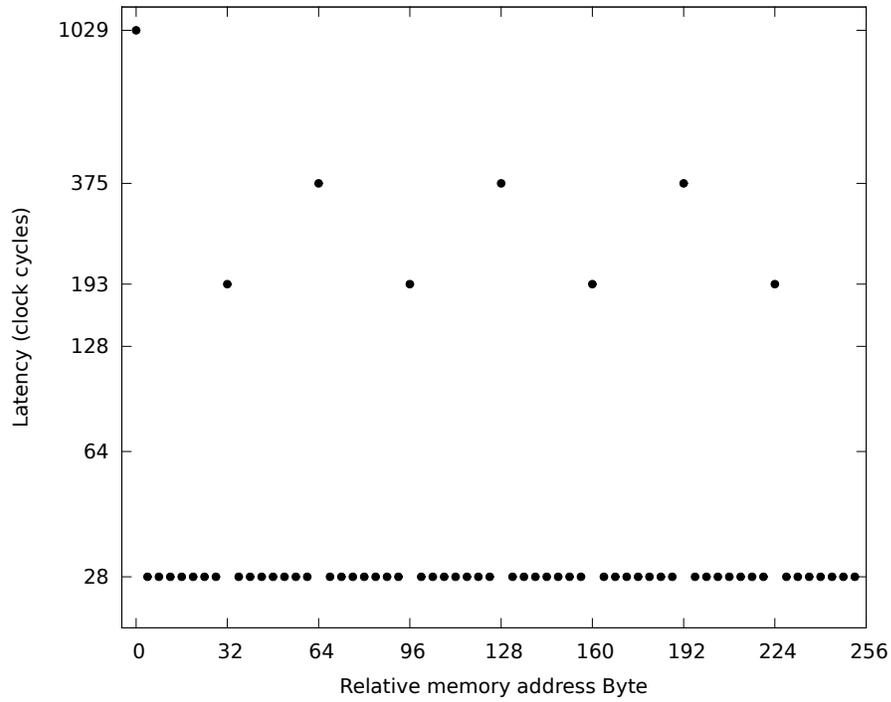

Figure 3.2: Global memory access latency, as per our measurements with the fine-grained p-chase method by Mei and Chu [12]. The 1029-cycle latency of the first access is the result of both L2 cache miss and TLB miss. The accesses to the following data, which are stored in the same L1 cache line, enjoy the very low, 28-cycle, L1 cache hit latency. The 193-cycle latency indicates an L1 cache miss and L2 cache hit, and the 375-cycle latency reflects an L2 cache miss latency with a TLB hit.

## 3.1   L1 data cache

### 3.1.1   Latency and bandwidth

As mentioned in the Volta architecture whitepaper [6], the combined L1 data cache and shared memory subsystem significantly reduces the cache hit latency and improves the bandwidth with respect to the performance on Pascal architecture.

We measured an average 28-cycle L1 cache hit latency on Volta (Figure 3.2) compared to an average latency of 84 cycles on Pascal.

Before Volta, Kepler was the most recent architecture to combine its L1

Table 3.2: L1 cache load throughput per SM.

|  | V100 | P100 | P4 | M60 | K80 | |
| --- | --- | --- | --- | --- | --- | --- |
| Measured Throughput | 108.3 | 31.3 | 15.7 | 15.7 | 68.6 | bytes/cycle |
| Theoretical upper bound | 256.0 | 128.0 | 128.0 | 256.0 | 256.0 | bytes/cycle |

cache and its shared memory. Kepler's L1 cache hit latency was 32 clock cycles.

We use the PTX-inlined benchmark in Listing 3.1 to measure the L1 cache load bandwidth. To increase the pressure of data load requests on the L1 cache, the benchmark scans an array with 64-bit elements; every warp accesses all the elements in the array.

As recorded in Table 3.2, the L1 data cache bandwidth we measure on the V100 GPU is 109.1 bytes per cycle per SM, which is more than 3 times higher than that of the P100 GPU (31.3 bytes per clock cycle per SM).

We calculate the theoretical throughput by multiplying the LSU count per SM by the number of bytes that each LSU can load per cycle per instruction.

Historically, architectures that employ an L1 cache combined with shared memory (Volta and Kepler) exhibit a higher L1 bandwidth than architectures where the L1 cached and the shared memory are separate (Pascal and Maxwell).

Listing 3.1: L1 data cache benchmark.

```
__global__ void l1_bw(uint32_t *startClk, uint32_t *stopClk,
                      double *dsink, uint32_t *posArray){
    // thread index
    uint32_t tid = threadIdx.x;

    // a register to avoid compiler optimization
    double sink = 0;

    // populate l1 cache to warm up
    for(uint32_t i = tid; i<L1_SIZE; i+=THREADS_NUM){
        double* ptr = posArray+i;
        asm volatile ("{\t\n"
          ".reg .f32 data;\n\t"
          "ld.global.ca.f64 data, [%1];\n\t"
          "add.f64 %0, data, %0;\n\t"
          "}" : "+d"(sink) : "l"(ptr) : "memory"
        );
    }

    // synchronize all threads
    asm volatile ("bar.sync 0;");

    // start timing
    uint32_t start = 0;
    asm volatile ("mov.u32 %0, %%clock;" : "=r"(start) :: "memory");

    // load data from l1 cache and accumulate
```

Table 3.3: Detectable L1 data cache size with the pointer-chase benchmark.

| Configured size of shared memory (KiB) | 0 | 64 | 96 |
|---|---|---|---|
| Expected size of L1 data cache (KiB) | 128 | 64 | 32 |
| Detected size of L1 data cache (KiB) | 121 | 57 | 25 |

```
    for(uint32_t i = 0; i<L1_SIZE; i+=THREADS_NUM){
        double* ptr = posArray+i;
        // every warp loads all data in l1 cache
        for(uint32_t j = 0; j<THREADS_NUM; j+=WARP_SIZE){
            uint32_t offset = (tid+j)%THREADS_NUM;
            asm volatile ("{\t\n"
                ".reg .f64 data;\n\t"
                "ld.global.ca.f64 data, [%1];\n\t"
                "add.f64 %0, data, %0;\n\t"
                "}" : "+d"(sink) : "l"(ptr+offset) : "memory"
            );
        }
    }

    // synchronize all threads
    asm volatile ("bar.sync 0;");

    // stop timing
    uint32_t stop = 0;
    asm volatile ("mov.u32 %0, %%clock;" : "=r"(stop) :: "memory");

    // write time and data back to memory
    startClk[tid] = start;
    stopClk[tid] = stop;
    dsink[tid] = sink;
}
```

### 3.1.2   Geometry

According to the V100 whitepaper [6], load/store operations can use a L1 data cache ranging from 32 KiB to 128 KiB in size.

Using Mei and Chu's fine-grained pointer-chase technique [12], our experiments were unable to detect the whole configured size and fell 7 KiB short of the nominal L1 data cache size (see Table 3.3).

We considered an experimental setup where the shared memory is configured to a size of 96-KiB. We then employed a benchmark that scans a variable length array **A** twice. As long as the size of array **A** exceeds 25 KiB, we detected cache misses.

At this time we are unable to explain this 7-KiB discrepancy. We conjecture it is the result of a newly applied replacement policy that we discuss below. We confirm that it is not associated to the ECC feature (error correction).

Table 3.1 describes the remainder of L1 data cache geometry as we discover it. The line size, load and update granularity of Volta's L1 data cache are the same as on the Pascal and Maxwell GPUs.

Table 3.4: L2 data cache load throughput.

|  | Volta V100 | Pascal P100 | Pascal P4 | Maxwell M60 | Kepler K80 |
|---|---|---|---|---|---|
| throughput (GB/s) | 2155 | 1624 | 979 | 446 | 339 |

Volta features an improved L1 cache replacement policy with respect to its predecessors. When the L1 data cache saturates, Volta replaces the same four cache lines first. These four cache lines are from different cache sets. Our benchmark reveals that Volta assigns L1 cache lines different priorities for preservation. The same four cache lines, from the 4 cache set, have the lowest preservation priority. These are replaced first and the remaining lines follow.

Compared to the commonly used LRU policy, this replacement policy better preserves large arrays from being evicted by sparse memory accesses.

Since shared memory is often used for storing large arrays, our findings agree with claim "*Volta narrows the gap between applications that explicitly manage shared memory and those that access data in device memory directly*" from the Volta whitepaper [6].

## 3.2   L2 cache

Volta employs an L2 cache that is unified for data, instructions and constant memory, as the previous GPU generations do. The L2 cache on the V100 GPU is an 16-way set-associative cache having a size of 6,144 KiB, a cache line of 64 B and an average latency of 193 clock cycles (Figure 3.2).

We measured the L2 cache load bandwidth on all the considered GPUs with the kernel in Listing 3.2, which loads data from L2 data cache and calculates a simple floating-point addition. Results are in Table 3.4.

Listing 3.2: L2 cache benchmark. Compared with the data accessing latency, the time consumed by floating-point accumulation is negligible.

```
__global__ void l2_bw(double *dsink, uint32_t *posArray){
    // block and thread index
    UINT tid = threadIdx.x;
    UINT bid = blockIdx.x;

    // a register to avoid compiler optimization
    double sink = 0;

    // load data from l2 cache and accumulate,
    // l2 cache is warmed up before the launch of this kernel.
    for(UINT i = 0; i<L2_SIZE; i+=THREADS_NUM){
```

```
    DTYPE* ptr = posArray+i;
    // every warp loads all data in l2 cache
    for(UINT j = 0; j<THREADS; j+=32){
      UINT offset = (tid+j)%THREADS;
      asm volatile ("{\t\n"
        ".reg .f64 data;\n\t"
        "ld.global.cg.f64 data, [%1];\n\t"
        "add.f64 %0, data, %0;\n\t"
        "}" : "+d"(sink) : "l"(ptr+offset) : "memory"
      );
    }
  }

  // store the result
  dsink[tid] = sink;
}
```

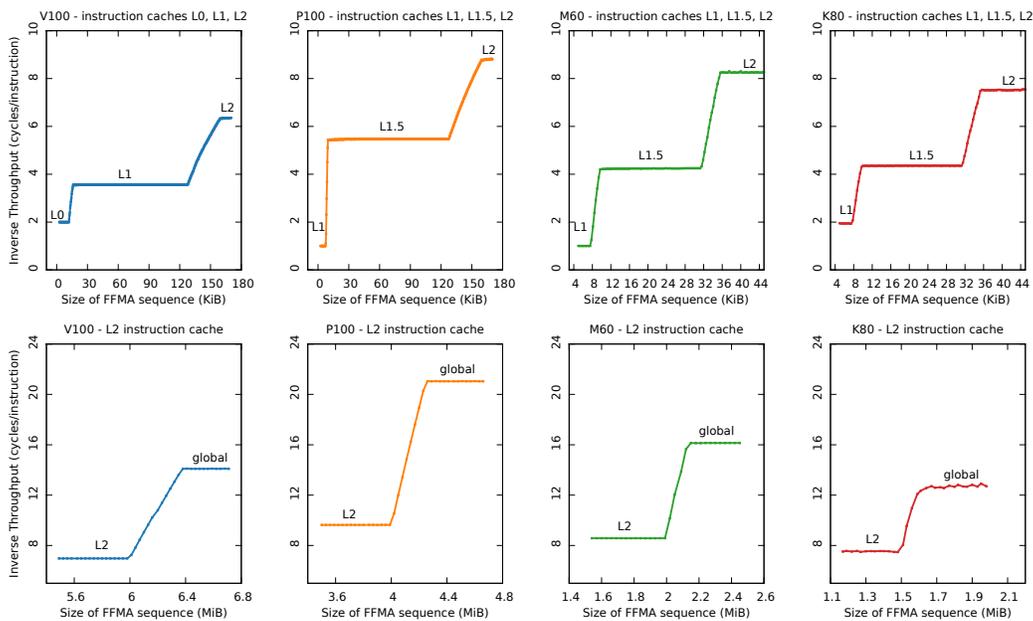

Figure 3.3: By charting the average inverse throughput experienced by instructions, we reveal the instruction cache hierarchy. Each plateau in the inverse throughput reveals a cache level. **Top charts**: the three plateaus correspond to cache hit on 3 instruction cache levels. **Bottom charts**: the benchmark saturates the L2 cache (of size 6, 4, 2, 1.5 MiB, respectively) and hits global memory. We collect the data by measuring the average clock cycles of a sequence of independent FFMA instructions. The FFMA instruction sequence is executed twice by one thread; our benchmark only records the clock cycle in the second execution to ensure instruction cache is warmed up. The end of each plateau indicates the size of a cache level. We chose FFMA instructions, operating only on registers, as to not cause pressure on the data memory hierarchy. We chose their operand registers so that each instruction experiences no register dependence with its neighbors.

## 3.3   Instruction caches

All NVIDIA GPU architectures we considered, including Volta, feature three levels of instruction caches.

However, we adopt a different naming scheme for instruction cache levels on Volta (L0, L1, L2) than on previous architectures (L1, L1.5, L2). This is consistent with with the V100 whitepaper [6] and with prior literature, which employ the two taxonomies.

Figure 3.3 charts the inverse throughput we measured on all the architectures considered. Volta enjoys better inverse throughput than its predecessors when accessing the second and third level of instruction caches.

To match the new partitioning method on Volta, the NVCC compiler tends to generate 2-cycle stalls between instructions, which causes the measured per-scheduler inverse throughput from L0 instruction cache to appear higher on Volta. However, since there are twice as many schedulers on every streaming processor (SM), the per-SM aggregate instruction load throughput from L0 cache doesn't decrease.

Across the different GPU architectures, levels in the instruction memory hierarchy are organized as follows:

- on Volta, each L0 cache is private to one scheduler/processing block;

- on all GPUs considered, each L1 instruction cache is private to an SMX;

- on Pascal, Maxwell and Kepler each L1.5 instruction cache is private to one SMX; the L1.5 instruction cache does not exist on Volta architecture.

- on all GPUs considered, the L2 cache is unified (i.e., it is used not only for instruction but for data as well) and it is shared across all SMXs.

By increasing the resolution of the benchmark, we determined the associativity of the L0 and L1 instruction cache. As shown in Figure 3.4, on Volta GPU, the L0 instruction cache is 3-way associative with 16 sets and 256 B cache line, the L1 instruction cache is 8-way associative with 32 sets and 512 B cache line.

We confirm that each L0 instruction cache is private to a processing block, each L1 instruction cache is private to an SM, and the L2 instruction cache is shared among all SMs (which is the unified L2 cache).

Figure 3.5 summarizes the findings that back up these claims. To prove that L0 instruction cache is private to every scheduler and that L1 is not, we use an experiment that invokes two warps (warp 0 and warp 1) on one Volta SM. Each warp records the inverse throughput of an FFMA sequence with fixed length. The fixed length FFMA instruction is warmed in instruction cache at the very beginning of the experiment. Before recording the latency, the benchmark thrashes the instruction cache on warp 0 with a variable length sequence. By increasing the size of thrashing sequence around the size of L0 cache gradually, we only observe that the recorded inverse throughput by warp 0 grows. We repeat the experiment by using longer thrashing (see the top-left part of Figure 3.5). We observe that the measured inverse throughput data from both warp 0 and warp 1 increase when the size of the thrashing sequence grows to the size of the L1 instruction cache (see the top-right part of Figure 3.5). Using similar experiments invoking two SMs (SM0 and SM1), we observe the recorded inverse throughput only grows on SM0 near the size of L1 instruction cache (see the bottom-left part of Figure 3.5); the inverse throughput grows on both SMs near the size of L2 instruction cache (see the bottom-right part of Figure 3.5).

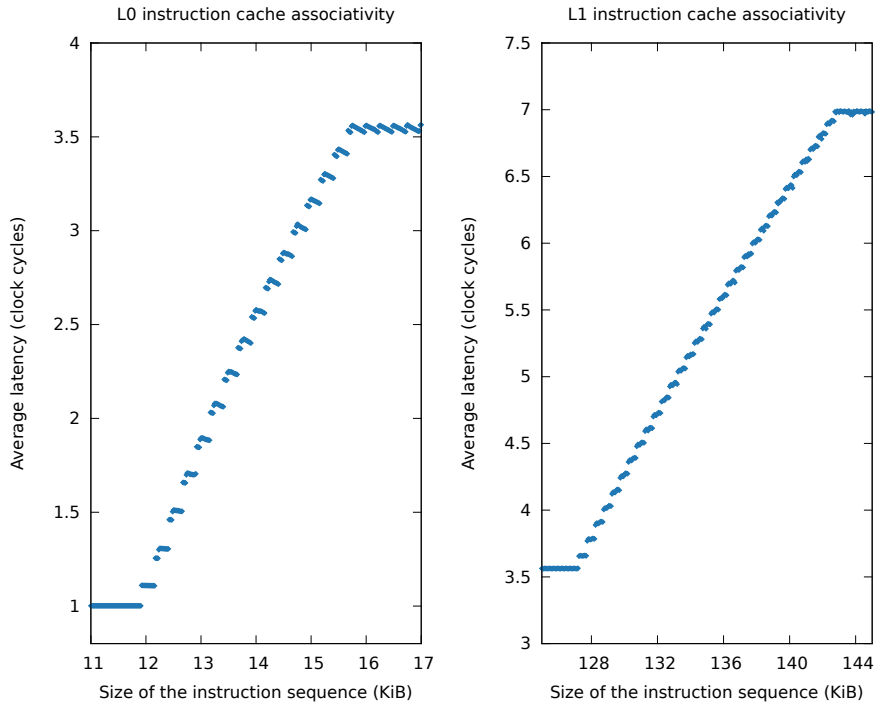

Figure 3.4: The measured inverse throughput of independent FFMA instructions. In the left (right) part, we can see 16 (32) clear stages when the tested instruction sequence saturates the L0 (L1) instruction cache.

In all experiments, we changed the size of the measured FFMA sequence to ensure our benchmark accesses the considered instruction cache level. For example, if the measured sequence is 1600 B, all the record instructions are initially cached in L0 instruction cache; if the measured sequence is 16000 B,
all the record instructions are initially cached in L1 instruction cache.

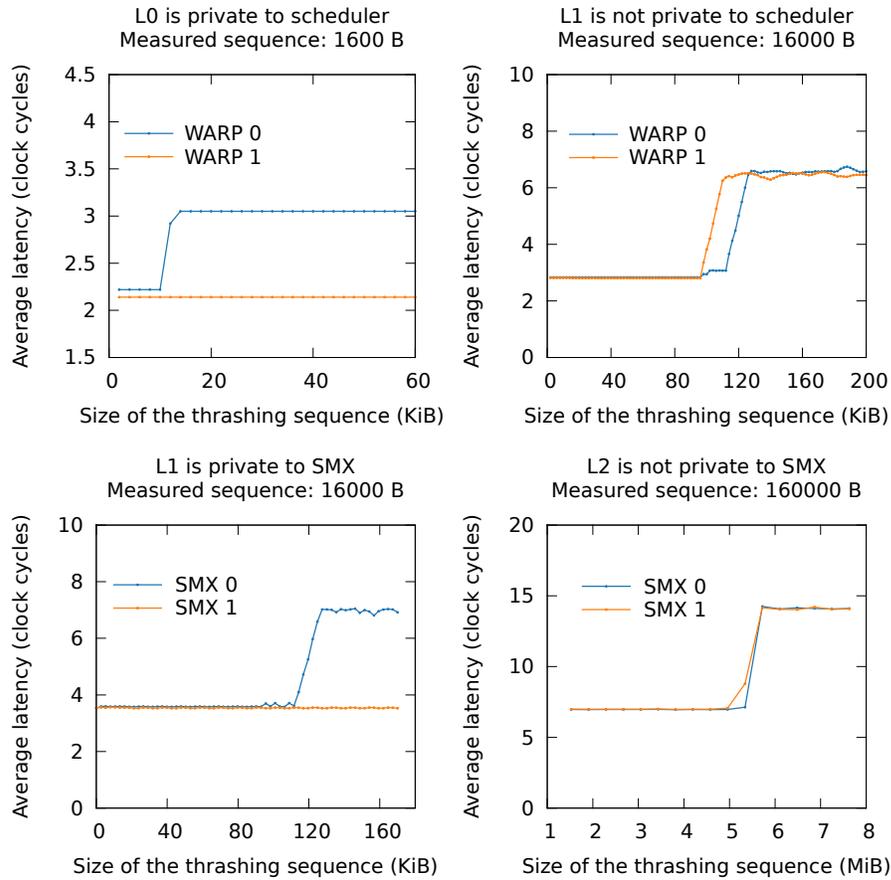

Figure 3.5:  The inverse throughput measured in the experiments to prove:
(1) L0 instruction cache is private to every scheduler (the top-left figure); (2)
L1 instruction cache is not private to every scheduler (the top-right figure);
(3) L1 instruction cache is private to every SM (the bottom-left figure); (4) L2
instruction cache is shared by all SMs (the bottom-right figure).

## 3.4   Constant memory and cache

Volta has three levels of constant cache memory, which have the geometry and
properties in Table 3.1 and latency as in Figure 3.7.

The constant memory hierarchy used in Volta did not change significantly

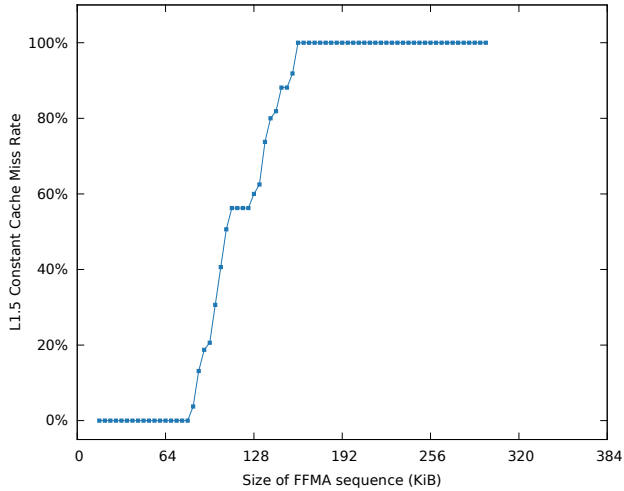

Figure 3.6: Measured L1.5 constant cache miss rate; Our benchmark thrashes L1 instruction cache with FFMA sequence between two constant array scans on L1.5 constant cache. We record the miss rate of the second scan.

from previous generations.

Across the GPU generations we considered, the following properties hold true:

- the L1 cache constant cache uses a non-LRU replacement policy;

- each SMX possesses two private levels of constant caches, which we denote as L1 and L1.5 constant cache (accesses to either of each level within an SMX do not affect the same cache levels on other SMXs);

- the L2 cache is the third level of constant cache. It is shared among all SMXs and is unified for instruction and data.

On Volta, the second levels of the constant and the instruction cache are the same. More precisely, the L1.5 constant cache and the L1 instruction cache coincide. Our experiment that fills the L1 instruction cache by running increasingly longer FFMA instruction sequences and, consequently, evicts the L1.5 constant cache proves this. As we vary the length of FFMA sequences, we see the miss rate in the L1.5 constant cache increase from 0% to 100% (Figure 3.6).

As in previous architectures, constant memory accesses on Volta support broadcasting (see Figure 3.7). When all threads within a warp access the same address, the constant memory sends data to all threads simultaneously. When threads visit diverging addresses, the accesses are serialized.

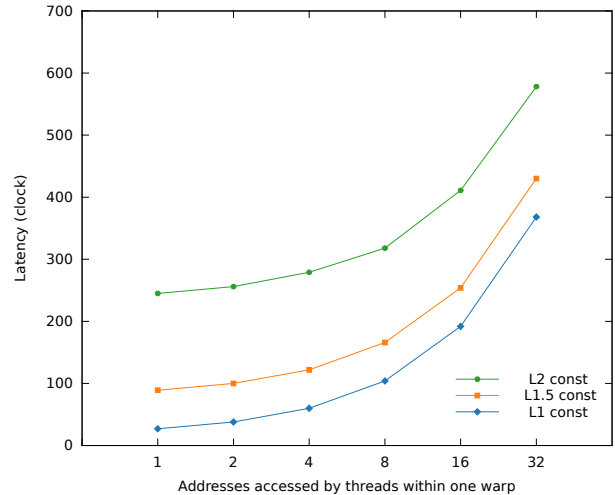

Figure 3.7: Latency of concurrent loads from constant memory within a warp depends on where the data is found in the cache hierarchy (L1, L1.5, or L2) and on the count of distinct locations referenced.  The hardware broadcasts accesses to the same location.

## 3.5   Register banks

The register file on Volta is divided into 2 banks and the width of each bank is 64 bits. This differs from the 4-bank, 32-bit wide design on previous architectures. The wider register banks make it easier for the compiler to generate conflict-free code. On Volta, the bank of a register is the register's index modulo 2.

Figure 3.8 displays the effect of register bank conflicts on the V100 GPU. We use sequences of identical FFMA instructions in which we vary one source register index (RX) to cause conflicts. Since the Volta GPU has 64-bit register banks, a conflict will only happen when all three 32-bit source registers in an "FFMA" instruction are in a same bank.  In every FFMA instruction in the benchmark of Figure 3.8, R97 and R99 are in bank 1; if RX also sits in bank 1, a conflict will occur.

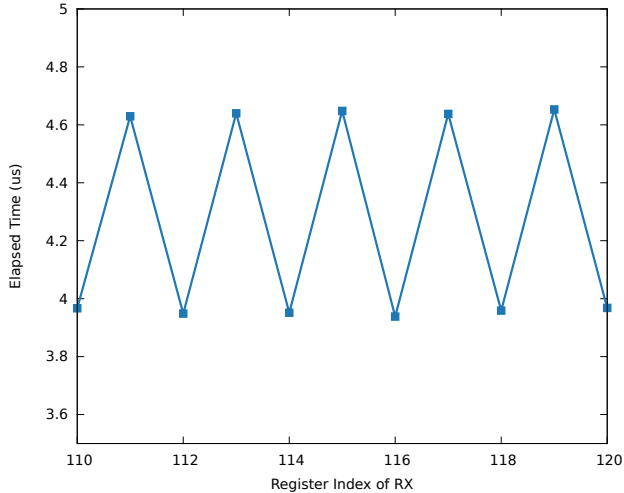

Figure 3.8: Elapsed time of identical FFMA instructions; we vary one source register index (RX) in an instruction sequence of "FFMA R6, R97, R99, RX" to cause conflicts. The choice of RX either causes one conflict or not.

## 3.6 Shared memory

The V100 GPU has up to 96 KiB of shared memory (configurable) with low latency and high memory bandwidth. In this section, we characterize shared memory performance under varying contention conditions.

**Latency**. Volta's shared memory has the lowest latency among the GPUs we examined (Figure 3.9). On all GPUs except for Kepler, the measured average access latency monotonically increases with the number of conflicts in a warp. Kepler is the only GPU adopting 64-bit (8-byte) wide banks. Kepler will not suffer latency degradation when two threads access the same shared memory bank. It will, however, suffer latency degradation when more than two threads are involved in the conflict.

**Bandwidth**. Due to the increased number of streaming multiprocessors, both the theoretical and measured shared memory bandwidth grow significantly in the Pascal and Volta GPUs (see Figure 3.10). We measured actual bandwidths using the `nvprof` profiler on a custom benchmark that every thread loads data from shared memory, performs a simple calculation and stores the results back to shared memory. The benchmark invokes as many threads and blocks as possible to provide enough pressure on LD/ST units.

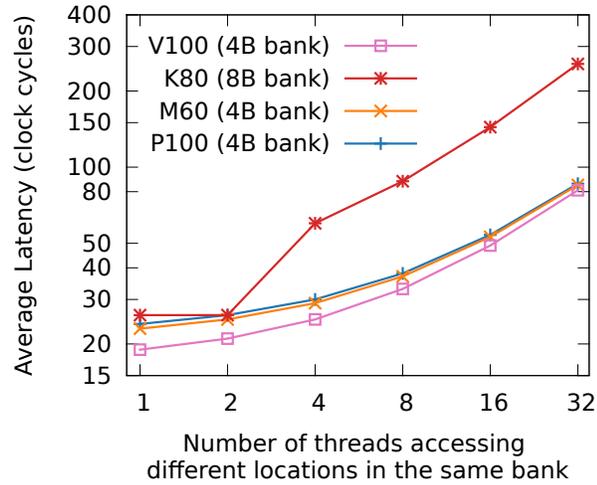

Figure 3.9: Shared memory latency increases under contention. Both axes use exponential scales. We used a stride factor multiplying the thread index as an offset to load data from shared memory. Each thread visits one 32-bit element and measures the average access latency

## 3.7  Global memory

We measured the actual global memory bandwidth and compared it against its theoretical limit for all the GPUs considered (Figure 3.11).

Thanks to their adoption of HBM2 memory, Volta and Pascal boards feature a significantly higher bandwidth than GPUs based on GDDR5 memory. The P100 outperforms GDDR-based GPUs boards but suffers from a larger gap between actual and theoretical performance.

Volta not only enjoys higher theoretical and actual bandwidth values than Pascal, it also enjoys a higher actual-to-theoretical bandwidth ratio (83.3% vs. 69.6%).

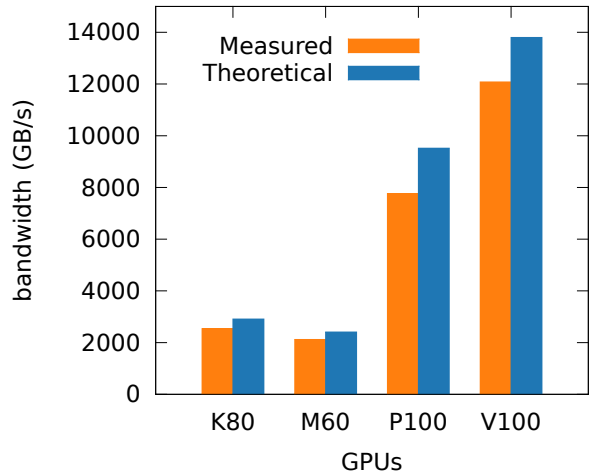

Figure 3.10: Theoretical and measured shared memory bandwidth on the considered GPUs. The theoretical limits are given by product $P \cdot B_s \cdot w_s \cdot f_g$. All these factors and their results can be found in Table 3.1.

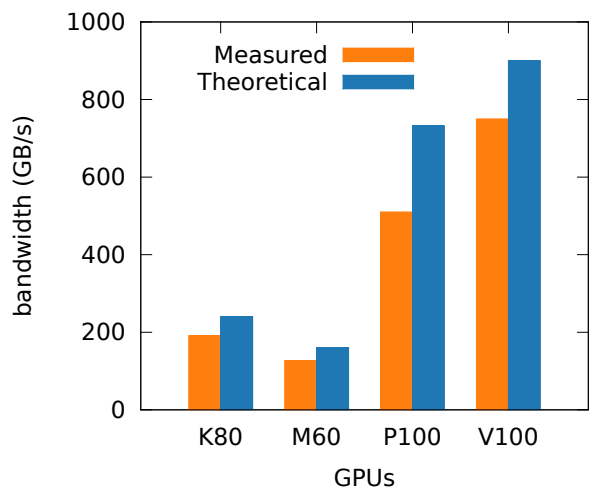

Figure 3.11: Theoretical and measured global memory bandwidth on all GPUs considered. We present theoretical bounds is from NVIDIA's whitepapers. Actual bandwiths are the results of our benchmark which loads data from a global array and stores it into another global array.

## 3.8   TLBs

On Volta and on all other architectures we examined:

- the L1 data cache is indexed by virtual addresses;

- the L2 data cache is indexed by physical addresses.

Because L2 is a physical cache, accesses to it involve the TLBs. We prove this claim by scanning a large array with L1 data cache enabled; we size the array to exceed the L1 TLB coverage, so that accesses in the benchmark would cause at least one level of TLB miss if L1 data cache were indexed by physical address. As expected, we saw no TLB misses in the second scan, as long as the stride is big enough to cache all accesses in L1 data cache. The same benchmark shows that addressing data in L2 data cache goes through the TLBs when the L1 data cache is disabled.

Figure 3.12 shows that, within the available global memory size, there are two levels of TLB on the Volta GPUs. The L1 TLB has 2 MiB page entries and 32 MiB coverage. The coverage of the L2 TLB is about 8192 MiB, which is larger than previous architectures.

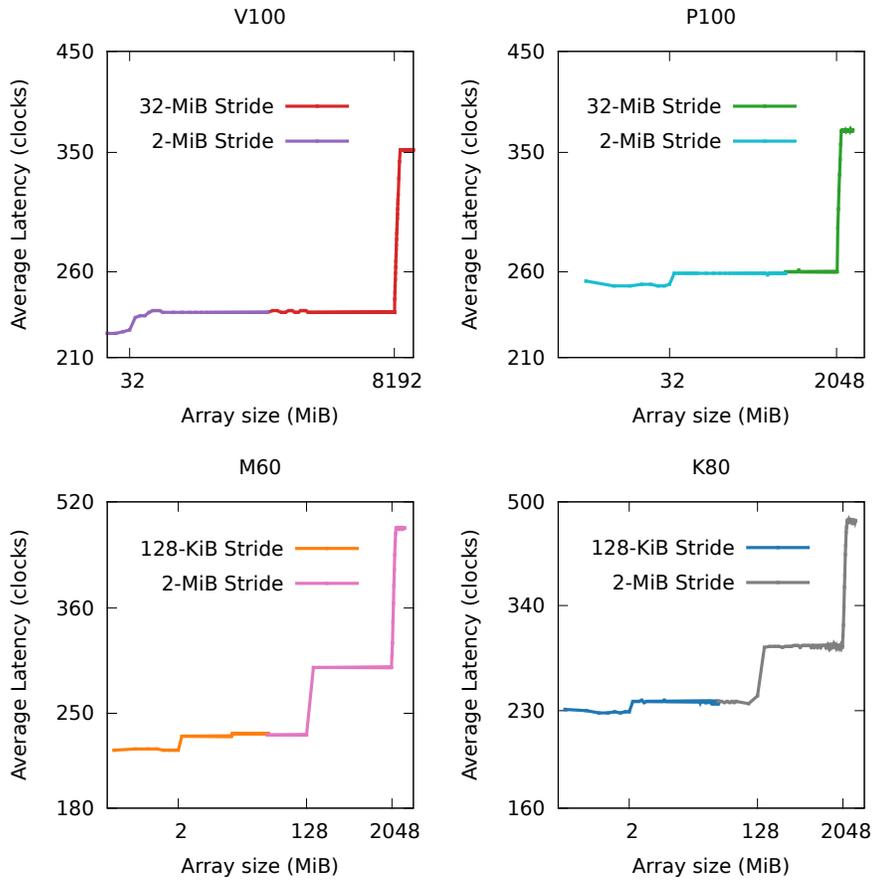

Figure 3.12: Global memory access latency seen by the pointer chase benchmark as it sweeps TLBs. The benchmarks perform a traditional pointer chase after a TLB warm-up scan, calculating the average global memory access latency with a stride of TLB page entry size.

# Chapter 4

# Instruction latency and throughput

In this chapter, we report the dependent instruction issue latency of native
Volta instructions and compare performance of atomics operations on all the
considered devices. We evaluate the floating-point performance in single,
double and half precision on our V100 GPU, and discuss the newly introduced
tensor core.

## 4.1 Native instructions

Comparing Volta against Pascal, we notice a generalized improvement in la-
tency.

We report the latency of common instructions on both Volta and Pascal in
Table 4.1.

Measuring instruction latency requires the use of benchmarks specifically
crafted for this purpose. To measure the latency of an instruction A, we add a
second instruction B that depends on A, then set the control word that regulate
A's execution:

- if A has fixed latency, we choose a B that consumes A's output. We
  decrease A's stall cycles in its control word, till A's result consumed by B
  is incorrect. The last stall value producing correct results is A's latency;

- if A has variable latency, we choose a B of known latency, then set control
  flags to create an artificial read/write dependency between A and B.
  We let the scheduler wait for the dependency, then measure the pair's
  cumulative latency with a bracket of `CS2R` instructions, and obtain A's
  latency by subtracting B's known one.

Table 4.1: Latency of frequently used instructions on Volta and Pascal.

| Architecture | Instructions | Latency (cycles) |
|---|---|---|
| Pascal | BFE, BFI, IADD, IADD32I, FADD, FMUL, FFMA, FMNMX, HADD2, HMUL2, HFMA2, IMNMX, ISCADD, LOP, LOP32I, LOP3, MOV, MOV32I, SEL, SHL, SHR, VADD, VABSDIFF, VMNMX, XMAD | 6 |
| | DADD, DMUL, DFMA, DMNMX | 8 |
| | FSET, DSET, DSETP, ISETP, FSETP | 12 |
| | POPC, FLO, MUFU, F2F, F2I, I2F, I2I | 14 |
| | IMUL, IMAD | ~86 |
| Volta | IADD3, SHF, LOP3, SEL, MOV, FADD, FFMA, FMUL, ISETP, FSET, FSETP | 4 |
| | IMAD, FMNMX, DSET, DSETP | 5 |
| | HADD2, HMUL2, HFMA2 | 6 |
| | DADD, DMUL, DFMA | 8 |
| | POPC | 10 |
| | FLO, BREV, MUFU | 14 |

As the Volta whitepaper [6] mentions, the dependent issue latency for core FMA math operations was reduced to 4 clock cycles on Volta.

On Pascal, most integer and single-precision instructions have a latency of 6 cycles; double-precision instructions have latency 8 cycles; more complex instructions, some of which run on the SFU, require 14 cycles.

On Maxwell and Pascal, instructions IMAD and IMUL have a long latency because they are emulated.

On Volta, most integer and single-precision instructions have a latency of 4 cycles; most double-precision instructions have latency 8 cycles; half-precision instructions have a latency of 6 cycles.

## 4.2 Atomic operations

Our measurements show that atomic operations on shared memory have a better latency on Volta than on Pascal and Maxwell (Table 4.2). This improvement is likely due to the improved shared memory latency on Volta architecture.

We see no obvious improvement in the latency of global memory atomics on Volta.

Among all the architectures considered, Kepler exhibits the highest latency gap between shared and global memory atomics. This is due to its lack of native support for shared memory atomics. Moreover, its emulated atomics degrade quickly under contention. Later architectures support atomics in hardware, and bring low latency and low contention penalties to atomics.

Table 4.2: Latency of atomic operations on shared and global memory, in clock cycles. We determine the latency of atomic instruction A by following it with a load instruction B, of known latency, that visits the same location. We deduce A's latency from that of pair (A,B) as done earlier.

| Contention | Shared memory | | | | | Global memory | | | | |
|---|---|---|---|---|---|---|---|---|---|---|
| | V100 | P100 | P4 | M60 | K80 | V100 | P100 | P4 | M60 | K80 |
| none | 6 | 15 | 16 | 17 | 93 | 36 | 26 | 30 | 24 | 29 |
| 2 threads | 7 | 17 | 18 | 19 | 214 | 31 | 31 | 50 | 26 | 69 |
| 4 threads | 11 | 19 | 25 | 25 | 460 | 32 | 48 | 50 | 41 | 96 |
| 8 threads | 18 | 30 | 30 | 31 | 952 | 41 | 48 | 51 | 41 | 152 |
| 16 threads | 24 | 46 | 46 | 47 | 1,936 | 58 | 50 | 51 | 46 | 264 |
| 32 threads | 66 | 78 | 78 | 79 | 4,257 | 76 | 50 | 51 | 46 | 488 |

Figure 4.1 reports the throughput measured on GPUs from Kepler to Volta in presence of contention, in four scenarios:

- *Scenario 1*, one block of 1,024 threads. Of these, $R$ threads access the same address, while the others access distinct, sequential addresses in global memory. 8 groups of threads access the same L2 cache line;

- *Scenario 2*, one block of 1,024 threads. Of these, $R$ threads access the same address, while the others access sequential L2 cache lines in global memory, with every group of threads accessing a single L2 cache line;

- *Scenario 3*, a variable number of blocks, of 1,024 threads each. All threads in all blocks access the same address; heavy contention exists among blocks;

- *Scenario 4*, a variable number of blocks, of 1,024 threads each. All threads within a block access the same address. Different blocks access distinct addresses; no contention exists among blocks.

The V100 GPU doesn't achieve the highest throughput in the scenarios with contention and the scenarios on single SM. The only scenario in which the V100 GPU achieves the best performance is on multiple SMs and without contention among SMs. From Maxwell to Pascal the aggregate throughput increase substantially.

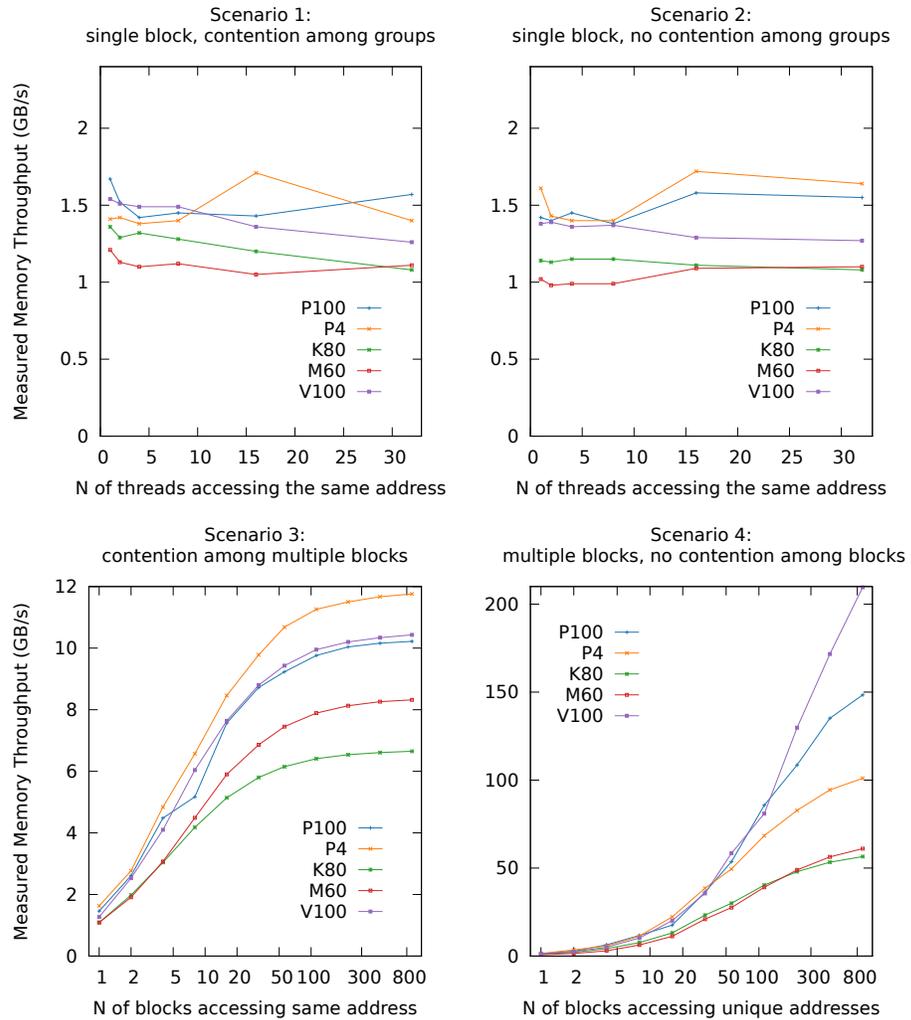

Figure 4.1: Throughput of *atomicAdd*() operations on global memory, measured in four contention scenarios.

## 4.3 Tensor cores

Volta introduces a new type of core, the *tensor core*. The tensor core takes half-precision floating point operands and provides higher throughput compared to the single precision cores.

Currently, CUDA programmers can only use warp-level primitive `wmma::mma_sync(acc_frag, a_frag, b_frag, acc_frag)` to perform $16 \times 16 \times 16$ half-precision matrix multiplication on tensor cores. In this section, we illustrate the detailed mechanism of computing on tensorcores by a column-major matrix multiplication $C = A \times B$ as our example. In the example, all three matrices are $16 \times 16$ in size. We discuss the generated instructions for tensor cores by the NVCC compiler, and analyze the runtime behavior of different threads. In our example, we use one warp to execute the considered multiplication. Elements in matrices $A$ and $B$ are half-precision floating-point number, and in matrix C are single-precision floating-point number.

Before invoking the matrix multiplication, programmers must load data from memory into registers with primitive `wmma::load_matrix_sync`, explicitly. The NVCC compiler translates that primitive into multiple memory load instructions. At run time, every thread loads 16 elements from matrix A and 16 elements from B.

We discovered the mapping between locations within each matrix. We also discovered the thread indices who loading the corresponding data into registers is fixed, and we report it in Figures 4.2 and 4.3 for matrices $A$ and $B$, respectively. At run time, two different threads load from one position in $A$ and $B$ into their own registers.

| 0 (0,8) | 32 (1,9) | 64 (2,10) | 96 (3,11) | 128 (0,8) | 160 (1,9) | 192 (2,10) | 224 (3,11) | 256 (0,8) | 288 (1,9) | 320 (2,10) | 352 (3,11) | 384 (0,8) | 416 (1,9) | 448 (2,10) | 480 (3,11) |
|---|---|---|---|---|---|---|---|---|---|---|---|---|---|---|---|
| 2 (0,8) | 34 (1,9) | 66 (2,10) | 98 (3,11) | 130 (0,8) | 162 (1,9) | 194 (2,10) | 226 (3,11) | 258 (0,8) | 290 (1,9) | 322 (2,10) | 354 (3,11) | 386 (0,8) | 418 (1,9) | 450 (2,10) | 482 (3,11) |
| 4 (0,8) | 36 (1,9) | 68 (2,10) | 100 (3,11) | 132 (0,8) | 164 (1,9) | 196 (2,10) | 228 (3,11) | 260 (0,8) | 292 (1,9) | 324 (2,10) | 356 (3,11) | 388 (0,8) | 420 (1,9) | 452 (2,10) | 484 (3,11) |
| 6 (0,8) | 38 (1,9) | 70 (2,10) | 102 (3,11) | 134 (0,8) | 166 (1,9) | 198 (2,10) | 230 (3,11) | 262 (0,8) | 294 (1,9) | 326 (2,10) | 358 (3,11) | 390 (0,8) | 422 (1,9) | 454 (2,10) | 486 (3,11) |
| 8 (16,24) | 40 (17,25) | 72 (18,26) | 104 (19,27) | 136 (16,24) | 168 (17,25) | 200 (18,26) | 232 (19,27) | 264 (16,24) | 296 (17,25) | 328 (18,26) | 360 (19,27) | 392 (16,24) | 424 (17,25) | 456 (18,26) | 488 (19,27) |
| 10 (16,24) | 42 (17,25) | 74 (18,26) | 106 (19,27) | 138 (16,24) | 170 (17,25) | 202 (18,26) | 234 (19,27) | 266 (16,24) | 298 (17,25) | 330 (18,26) | 362 (19,27) | 394 (16,24) | 426 (17,25) | 458 (18,26) | 490 (19,27) |
| 12 (16,24) | 44 (17,25) | 76 (18,26) | 108 (19,27) | 140 (16,24) | 172 (17,25) | 204 (18,26) | 236 (19,27) | 268 (16,24) | 300 (17,25) | 332 (18,26) | 364 (19,27) | 396 (16,24) | 428 (17,25) | 460 (18,26) | 492 (19,27) |
| 14 (16,24) | 46 (17,25) | 78 (18,26) | 110 (19,27) | 142 (16,24) | 174 (17,25) | 206 (18,26) | 238 (19,27) | 270 (16,24) | 302 (17,25) | 334 (18,26) | 366 (19,27) | 398 (16,24) | 430 (17,25) | 462 (18,26) | 494 (19,27) |
| 16 (4,12) | 48 (5,13) | 80 (6,14) | 112 (7,15) | 144 (4,12) | 176 (5,13) | 208 (6,14) | 240 (7,15) | 272 (4,12) | 304 (5,13) | 336 (6,14) | 368 (7,15) | 400 (4,12) | 432 (5,13) | 464 (6,14) | 496 (7,15) |
| 18 (4,12) | 50 (5,13) | 82 (6,14) | 114 (7,15) | 146 (4,12) | 178 (5,13) | 210 (6,14) | 242 (7,15) | 274 (4,12) | 306 (5,13) | 338 (6,14) | 370 (7,15) | 402 (4,12) | 434 (5,13) | 466 (6,14) | 498 (7,15) |
| 20 (4,12) | 52 (5,13) | 84 (6,14) | 116 (7,15) | 148 (4,12) | 180 (5,13) | 212 (6,14) | 244 (7,15) | 276 (4,12) | 308 (5,13) | 340 (6,14) | 372 (7,15) | 404 (4,12) | 436 (5,13) | 468 (6,14) | 500 (7,15) |
| 22 (4,12) | 54 (5,13) | 86 (6,14) | 118 (7,15) | 150 (4,12) | 182 (5,13) | 214 (6,14) | 246 (7,15) | 278 (4,12) | 310 (5,13) | 342 (6,14) | 374 (7,15) | 406 (4,12) | 438 (5,13) | 470 (6,14) | 502 (7,15) |
| 24 (20,28) | 56 (21,29) | 88 (22,30) | 120 (23,31) | 152 (20,28) | 184 (21,29) | 216 (22,30) | 248 (23,31) | 280 (20,28) | 312 (21,29) | 344 (22,30) | 376 (23,31) | 408 (20,28) | 440 (21,29) | 472 (22,30) | 504 (23,31) |
| 26 (20,28) | 58 (21,29) | 90 (22,30) | 122 (23,31) | 154 (20,28) | 186 (21,29) | 218 (22,30) | 250 (23,31) | 282 (20,28) | 314 (21,29) | 346 (22,30) | 378 (23,31) | 410 (20,28) | 442 (21,29) | 474 (22,30) | 506 (23,31) |
| 28 (20,28) | 60 (21,29) | 92 (22,30) | 124 (23,31) | 156 (20,28) | 188 (21,29) | 220 (22,30) | 252 (23,31) | 284 (20,28) | 316 (21,29) | 348 (22,30) | 380 (23,31) | 412 (20,28) | 444 (21,29) | 476 (22,30) | 508 (23,31) |
| 30 (20,28) | 62 (21,29) | 94 (22,30) | 126 (23,31) | 158 (20,28) | 190 (21,29) | 222 (22,30) | 254 (23,31) | 286 (20,28) | 318 (21,29) | 350 (22,30) | 382 (23,31) | 414 (20,28) | 446 (21,29) | 478 (22,30) | 510 (23,31) |

Figure 4.2: Mapping between positions in matrix A (column major) and thread indices in loading data. Every number before a parentheses is the relative address in matrix A in byte, every number within a parentheses is the thread index to load data from the corresponding position.

| 0 (0,4) | 32 (1,5) | 64 (2,6) | 96 (3,7) | 128 (16,20) | 160 (17,21) | 192 (18,22) | 224 (19,23) | 256 (8,12) | 288 (9,13) | 320 (10,14) | 352 (11,15) | 384 (24,28) | 416 (25,29) | 448 (26,30) | 480 (27,31) |
|---|---|---|---|---|---|---|---|---|---|---|---|---|---|---|---|
| 2 (0,4) | 34 (1,5) | 66 (2,6) | 98 (3,7) | 130 (16,20) | 162 (17,21) | 194 (18,22) | 226 (19,23) | 258 (8,12) | 290 (9,13) | 322 (10,14) | 354 (11,15) | 386 (24,28) | 418 (25,29) | 450 (26,30) | 482 (27,31) |
| 4 (0,4) | 36 (1,5) | 68 (2,6) | 100 (3,7) | 132 (16,20) | 164 (17,21) | 196 (18,22) | 228 (19,23) | 260 (8,12) | 292 (9,13) | 324 (10,14) | 356 (11,15) | 388 (24,28) | 420 (25,29) | 452 (26,30) | 484 (27,31) |
| 6 (0,4) | 38 (1,5) | 70 (2,6) | 102 (3,7) | 134 (16,20) | 166 (17,21) | 198 (18,22) | 230 (19,23) | 262 (8,12) | 294 (9,13) | 326 (10,14) | 358 (11,15) | 390 (24,28) | 422 (25,29) | 454 (26,30) | 486 (27,31) |
| 8 (0,4) | 40 (1,5) | 72 (2,6) | 104 (3,7) | 136 (16,20) | 168 (17,21) | 200 (18,22) | 232 (19,23) | 264 (8,12) | 296 (9,13) | 328 (10,14) | 360 (11,15) | 392 (24,28) | 424 (25,29) | 456 (26,30) | 488 (27,31) |
| 10 (0,4) | 42 (1,5) | 74 (2,6) | 106 (3,7) | 138 (16,20) | 170 (17,21) | 202 (18,22) | 234 (19,23) | 266 (8,12) | 298 (9,13) | 330 (10,14) | 362 (11,15) | 394 (24,28) | 426 (25,29) | 458 (26,30) | 490 (27,31) |
| 12 (0,4) | 44 (1,5) | 76 (2,6) | 108 (3,7) | 140 (16,20) | 172 (17,21) | 204 (18,22) | 236 (19,23) | 268 (8,12) | 300 (9,13) | 332 (10,14) | 364 (11,15) | 396 (24,28) | 428 (25,29) | 460 (26,30) | 492 (27,31) |
| 14 (0,4) | 46 (1,5) | 78 (2,6) | 110 (3,7) | 142 (16,20) | 174 (17,21) | 206 (18,22) | 238 (19,23) | 270 (8,12) | 302 (9,13) | 334 (10,14) | 366 (11,15) | 398 (24,28) | 430 (25,29) | 462 (26,30) | 494 (27,31) |
| 16 (0,4) | 48 (1,5) | 80 (2,6) | 112 (3,7) | 144 (16,20) | 176 (17,21) | 208 (18,22) | 240 (19,23) | 272 (8,12) | 304 (9,13) | 336 (10,14) | 368 (11,15) | 400 (24,28) | 432 (25,29) | 464 (26,30) | 496 (27,31) |
| 18 (0,4) | 50 (1,5) | 82 (2,6) | 114 (3,7) | 146 (16,20) | 178 (17,21) | 210 (18,22) | 242 (19,23) | 274 (8,12) | 306 (9,13) | 338 (10,14) | 370 (11,15) | 402 (24,28) | 434 (25,29) | 466 (26,30) | 498 (27,31) |
| 20 (0,4) | 52 (1,5) | 84 (2,6) | 116 (3,7) | 148 (16,20) | 180 (17,21) | 212 (18,22) | 244 (19,23) | 276 (8,12) | 308 (9,13) | 340 (10,14) | 372 (11,15) | 404 (24,28) | 436 (25,29) | 468 (26,30) | 500 (27,31) |
| 22 (0,4) | 54 (1,5) | 86 (2,6) | 118 (3,7) | 150 (16,20) | 182 (17,21) | 214 (18,22) | 246 (19,23) | 278 (8,12) | 310 (9,13) | 342 (10,14) | 374 (11,15) | 406 (24,28) | 438 (25,29) | 470 (26,30) | 502 (27,31) |
| 24 (0,4) | 56 (1,5) | 88 (2,6) | 120 (3,7) | 152 (16,20) | 184 (17,21) | 216 (18,22) | 248 (19,23) | 280 (8,12) | 312 (9,13) | 344 (10,14) | 376 (11,15) | 408 (24,28) | 440 (25,29) | 472 (26,30) | 504 (27,31) |
| 26 (0,4) | 58 (1,5) | 90 (2,6) | 122 (3,7) | 154 (16,20) | 186 (17,21) | 218 (18,22) | 250 (19,23) | 282 (8,12) | 314 (9,13) | 346 (10,14) | 378 (11,15) | 410 (24,28) | 442 (25,29) | 474 (26,30) | 506 (27,31) |
| 28 (0,4) | 60 (1,5) | 92 (2,6) | 124 (3,7) | 156 (16,20) | 188 (17,21) | 220 (18,22) | 252 (19,23) | 284 (8,12) | 316 (9,13) | 348 (10,14) | 380 (11,15) | 412 (24,28) | 444 (25,29) | 476 (26,30) | 508 (27,31) |
| 30 (0,4) | 62 (1,5) | 94 (2,6) | 126 (3,7) | 158 (16,20) | 190 (17,21) | 222 (18,22) | 254 (19,23) | 286 (8,12) | 318 (9,13) | 350 (10,14) | 382 (11,15) | 414 (24,28) | 446 (25,29) | 478 (26,30) | 510 (27,31) |

Figure 4.3: Mapping between positions in matrix B (column major) and thread indices to load data. Every number before a parenthesis is the relative address in matrix A in byte, every number within a parenthesis is the thread index to load data from the corresponding position.

Listing 4.1: Assembly code generated for a `mma_sync` primitive by nvcc from CUDA 9.0.

```
; set 0:
HMMA.884.F32.F32.STEP0 R8, R26.reuse.T, R16.reuse.T, R8;
HMMA.884.F32.F32.STEP1 R10, R26.reuse.T, R16.reuse.T, R10;
HMMA.884.F32.F32.STEP2 R4, R26.reuse.T, R16.reuse.T, R4;
HMMA.884.F32.F32.STEP3 R6, R26.T, R16.T, R6;

; set 1:
HMMA.884.F32.F32.STEP0 R8, R20.reuse.T, R18.reuse.T, R8;
HMMA.884.F32.F32.STEP1 R10, R20.reuse.T, R18.reuse.T, R10;
HMMA.884.F32.F32.STEP2 R4, R20.reuse.T, R18.reuse.T, R4;
HMMA.884.F32.F32.STEP3 R6, R20.T, R18.T, R6;

; set 2:
HMMA.884.F32.F32.STEP0 R8, R22.reuse.T, R12.reuse.T, R8;
HMMA.884.F32.F32.STEP1 R10, R22.reuse.T, R12.reuse.T, R10;
HMMA.884.F32.F32.STEP2 R4, R22.reuse.T, R12.reuse.T, R4;
HMMA.884.F32.F32.STEP3 R6, R22.T, R12.T, R6;

; set 3:
HMMA.884.F32.F32.STEP0 R8, R2.reuse.T, R14.reuse.T, R8;
HMMA.884.F32.F32.STEP1 R10, R2.reuse.T, R14.reuse.T, R10;
HMMA.884.F32.F32.STEP2 R4, R2.reuse.T, R14.reuse.T, R4;
HMMA.884.F32.F32.STEP3 R6, R2.T, R14.T, R6;
```

After loading data from memory into registers, programmers must use `wmma::mma_sync` to compute on tensor cores. The NVCC compiler translates the single `mma_sync` primitive into 4 sets of HMMA instructions, as in Listing 4.1. Each set consists of 4 HMMA instructions with flags from "STEP0" to "STEP3". The target registers are the same among all 4 instruction sets. Within every set, flags "STEP0" to "STEP3" correspond to computing different positions of $C$ (see Figure 4.4).

At run time, Volta divides the 32 threads of a warp into 8 groups ($group\_id = thread\_id/4$) and shares register values among threads within the same group, and among groups. Every thread group computes 4×8=32 elements into matrix $C$. Figure 4.5 shows the mapping between positions in matrix $C$ and group indices. For one thread group, all HMMA instruction sets contribute to the same positions in the matrix $C$, and they each accumulate and multiply elements from different parts of matrix $A$ and $B$ (see Figure 4.6).

Finally, programmers must use primitive `wmma::store_matrix_sync` to write back results from the registers distributed across the threads, back into matrix $C$. Figure 4.7 displays the mapping between positions in matrix $C$ and thread indices in our example.

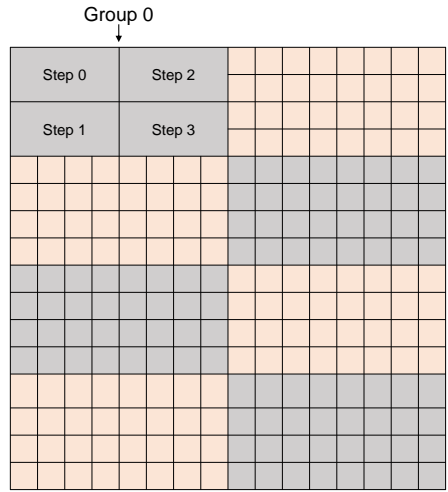

Figure 4.4: 4 steps of HMMA instructions within one set compute different elements in matrix $C$.

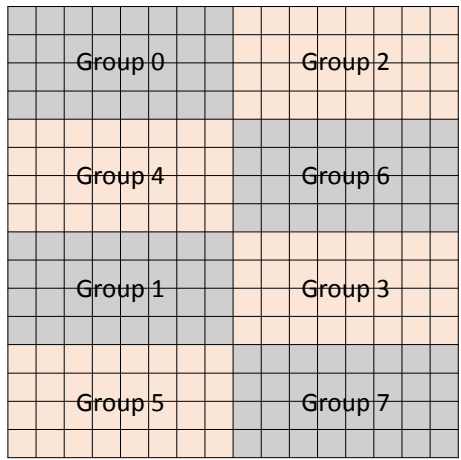

Figure 4.5: Mapping between positions in matrix $C$ and thread group indices.

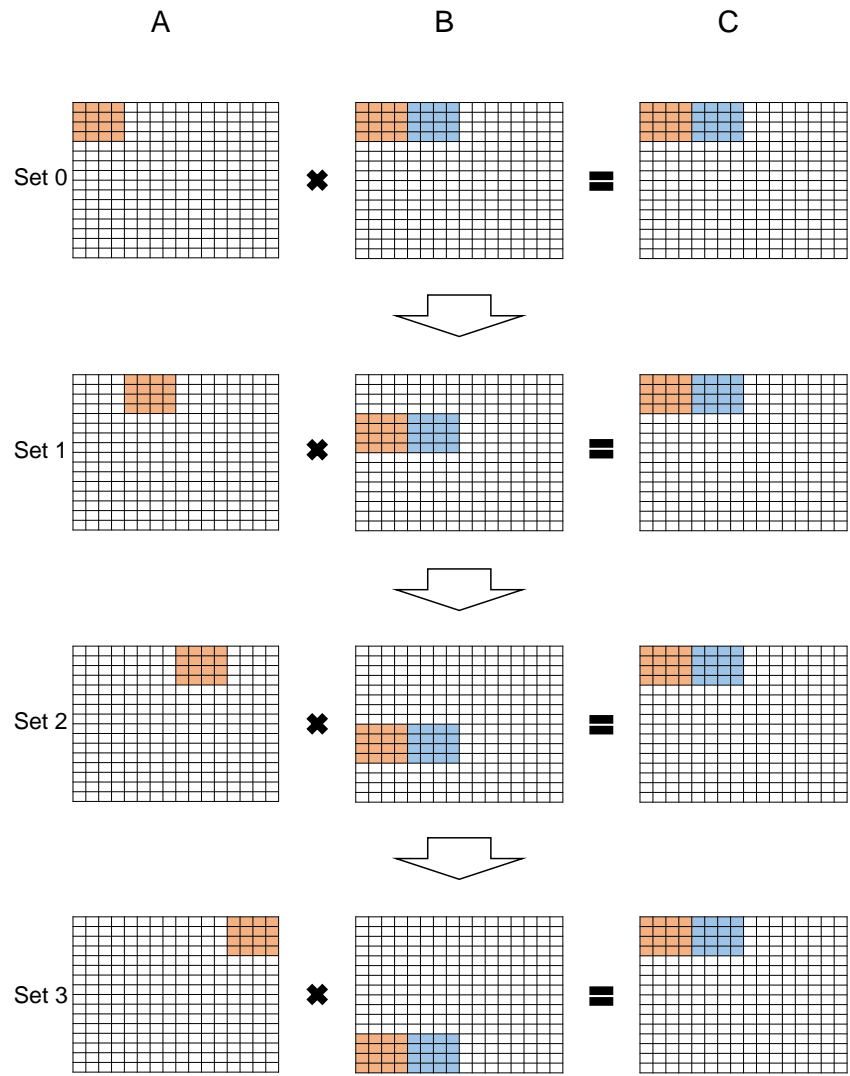

Figure 4.6: Four sets of HMMA instructions complete 4×8 results in matrix $C$ within thread group 0. Different sets use different elements in $A$ and $B$. The instructions in set 0 execute first, then the instructions in set 1, set 2 and set 3. This way, the 16 HMMA instructions can correctly compute the 4×8 elements in matrix $C$.

| | | | | | | | | | | | | | | | |
|---|---|---|---|---|---|---|---|---|---|---|---|---|---|---|---|
| 0 (0) | 64 (0) | 128 (2) | 192 (2) | 256 (0) | 320 (0) | 384 (2) | 448 (2) | 512 (8) | 576 (8) | 640 (10) | 704 (10) | 768 (8) | 832 (8) | 896 (10) | 960 (10) |
| 4 (1) | 68 (1) | 132 (3) | 196 (3) | 260 (1) | 324 (1) | 388 (3) | 452 (3) | 516 (9) | 580 (9) | 644 (11) | 708 (11) | 772 (9) | 836 (9) | 900 (11) | 964 (11) |
| 8 (0) | 72 (0) | 136 (2) | 200 (2) | 264 (0) | 328 (0) | 392 (2) | 456 (2) | 520 (8) | 584 (8) | 648 (10) | 712 (10) | 776 (8) | 840 (8) | 904 (10) | 968 (10) |
| 12 (1) | 76 (1) | 140 (3) | 204 (3) | 268 (1) | 332 (1) | 396 (3) | 460 (3) | 524 (9) | 588 (9) | 652 (11) | 716 (11) | 780 (9) | 844 (9) | 908 (11) | 972 (11) |
| 16 (16) | 80 (16) | 144 (18) | 208 (18) | 272 (16) | 336 (16) | 400 (18) | 464 (18) | 528 (24) | 592 (24) | 656 (26) | 720 (26) | 784 (24) | 848 (24) | 912 (26) | 976 (26) |
| 20 (17) | 84 (17) | 148 (19) | 212 (19) | 276 (17) | 340 (17) | 404 (19) | 468 (19) | 532 (25) | 596 (25) | 660 (27) | 724 (27) | 788 (25) | 852 (25) | 916 (27) | 980 (27) |
| 24 (16) | 88 (16) | 152 (18) | 216 (18) | 280 (16) | 344 (16) | 408 (18) | 472 (18) | 536 (24) | 600 (24) | 664 (26) | 728 (26) | 792 (24) | 856 (24) | 920 (26) | 984 (26) |
| 28 (17) | 92 (17) | 156 (19) | 220 (19) | 284 (17) | 348 (17) | 412 (19) | 476 (19) | 540 (25) | 604 (25) | 668 (27) | 732 (27) | 796 (25) | 860 (25) | 924 (27) | 988 (27) |
| 32 (4) | 96 (4) | 160 (6) | 224 (6) | 288 (4) | 352 (4) | 416 (6) | 480 (6) | 544 (12) | 608 (12) | 672 (14) | 736 (14) | 800 (12) | 864 (12) | 928 (14) | 992 (14) |
| 36 (5) | 100 (5) | 164 (7) | 228 (7) | 292 (5) | 356 (5) | 420 (7) | 484 (7) | 548 (13) | 612 (13) | 676 (15) | 740 (15) | 804 (13) | 868 (13) | 932 (15) | 996 (15) |
| 40 (4) | 104 (4) | 168 (6) | 232 (6) | 296 (4) | 360 (4) | 424 (6) | 488 (6) | 552 (12) | 616 (12) | 680 (14) | 744 (14) | 808 (12) | 872 (12) | 936 (14) | 1000 (14) |
| 44 (5) | 108 (5) | 172 (7) | 236 (7) | 300 (5) | 364 (5) | 428 (7) | 492 (7) | 556 (13) | 620 (13) | 684 (15) | 748 (15) | 812 (13) | 876 (13) | 940 (15) | 1004 (15) |
| 48 (20) | 112 (20) | 176 (22) | 240 (22) | 304 (20) | 368 (20) | 432 (22) | 496 (22) | 560 (28) | 624 (28) | 688 (30) | 752 (30) | 816 (28) | 880 (28) | 944 (30) | 1008 (30) |
| 52 (21) | 116 (21) | 180 (23) | 244 (23) | 308 (21) | 372 (21) | 436 (23) | 500 (23) | 564 (29) | 628 (29) | 692 (31) | 756 (31) | 820 (29) | 884 (29) | 948 (31) | 1012 (31) |
| 56 (20) | 120 (20) | 184 (22) | 248 (22) | 312 (20) | 376 (20) | 440 (22) | 504 (22) | 568 (28) | 632 (28) | 696 (30) | 760 (30) | 824 (28) | 888 (28) | 952 (30) | 1016 (30) |
| 60 (21) | 124 (21) | 188 (23) | 252 (23) | 316 (21) | 380 (21) | 444 (23) | 508 (23) | 572 (29) | 636 (29) | 700 (31) | 764 (31) | 828 (29) | 892 (29) | 956 (31) | 1020 (31) |

Figure 4.7: Mapping between positions in matrix $C$ (column major) and thread indices that store data back into memory. Every number before a parentheses is the relative address in matrix C in byte, every number within parentheses is the thread index to store data to the corresponding position.

## 4.4   Floating-point performance

We evaluated the floating-point performance of a V100 GPU with the cuBLAS library shipped with CUDA 9.0. We report the arithmetic throughput (in TFLOPS) of matrix multiplication operations measured on a V100 running at 1,380 MHz at different levels of precision.

In single- and double- precision floating point operations, benchmarks achieve near-peak performance. In half precision, using the tensor cores, our measured performance reaches 90.2% of the theoretical throughput (Figure 4.8).

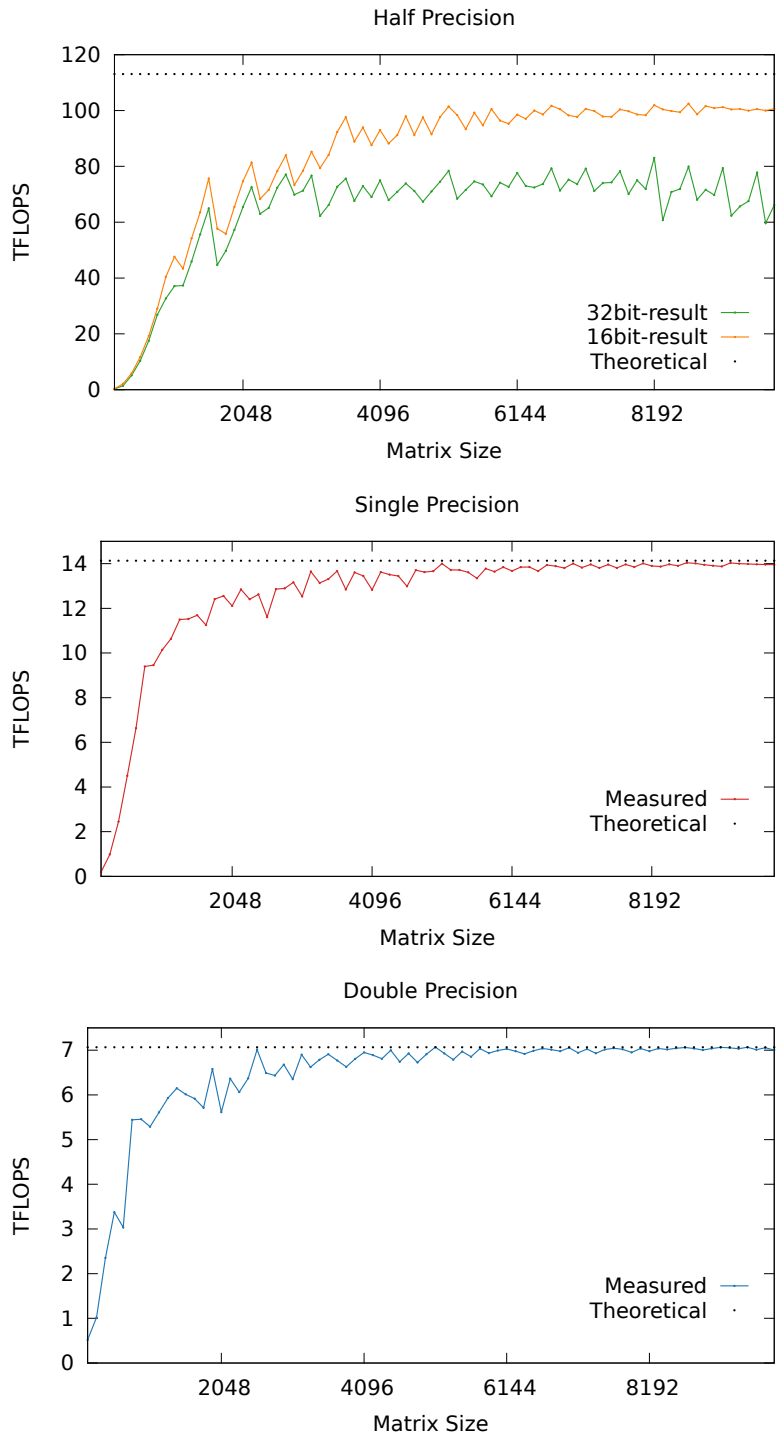

Figure 4.8: Floating-point performance of cuBLAS matrix multiplication on a V100 GPU running at 1,380 MHz.

# Chapter 5

# Volta boards with NVIDIA NVLink support

At the time of this writing, V100 GPUs are available in two forms:

- V100 boards in SXM2 form factor, supporting the NVLink 2.0 interconnect;

- V100 boards in PCIe form factor, supporting PCIe Gen 3 interconnect.

All data we reported in the previous chapters refer to the PCIe version of all boards, including Volta and Pascal.

In this chapter we provide data for the V100 NVLink GPU and compare its performance with the V100 PCIe GPU and the Pascal P100 NVLink GPU. We benchmark all GPUs at the respective highest peak graphic clock frequency:

| Device | Peak graphics clock frequency |
|---|---|
| V100 PCIe | 1,380 MHz |
| V100 NVLink | 1,530 MHz |
| P100 PCIe | 1,328 MHz |

## 5.1   Peer-to-peer and host communication

The NVLink 2.0 interconnect allows for up to 6 sub-links to each V100 GPU. Since the theoretical data rate per sub-link per direction is 25 GB/s, a V100 GPU can enjoy up to 300 GB/s interconnect bandwidth [19].

Table 5.1 reports peer-to-peer communication latencies and bandwidths measured among P100 NVLink GPUs (NVLink 1.0), V100 PCIe GPUs (PCIe Gen 3) and V100 NVLink GPUs (NVLink 2.0).

V100 NVLink GPUs enjoy significantly higher peer-to-peer bandwidth than P100 NVLink boards. NVLink GPUs, as expected, enjoy significantly higher peer-to-peer bandwidth than PCI boards.

Latencies are substantially similar across GPUs of different generations and type of interconnect.

Table 5.1: Communication latency and bandwidth we measured on V100 and P100 GPUs. For the GPUs supporting NVLink, we measure performance with peer-to-peer communication enabled, and only report the performance of GPUs connected by two sub-links. NVLink 1.0 only allows one GPU to connect with 4 sub-links.

|                              | V100 PCIe | P100 NVLink | V100 NVLink |      |
| ---------------------------- | --------- | ----------- | ----------- | ---- |
| Unidirection bandwidth       | 10.63     | 36.72       | 47.99       | GB/s |
| Latency within same PCIe switch | 7.21   | 9.47        | 8.55        | us   |

Since our GPUs are all connected to one PCIe Switch, which talks with one CPU through x16 Gen 3 PCI express, the host-to-device and device-to-host bandwidth data are same when only one GPU is communicating to its closest CPU on all hosts where we performed measurements (see Table 5.2).

Table 5.2: Bandwidth between host and device, measured on V100 and P100 GPUs.

|                        | V100 PCIe | P100 NVLink | V100 NVLink |      |
| ---------------------- | --------- | ----------- | ----------- | ---- |
| Host-to-device bandwidth | 12,152.4 | 12,135.9   | 12,147.8    | MB/s |
| Device-to-host bandwidth | 12,881.1 | 12,845.9   | 12,858.0    | MB/s |

## 5.2   Arithmetic throughput and memory bandwidth

The higher peak graphics clock on the NVLink version of the V100 board (1,530 vs. 1,380 MHz) explains why it enjoys higher performance than its PCIe version in the following aspects:

- floating-point throughput;

- L1 cache bandwidth;

- L2 cache bandwidth;

- shared memory bandwidth;

- atomic throughput when tested on multiple blocks.

We compare the floating-point throughput and memory bandwidth data in Table 5.3.

We repeat (on the NVLink V100 GPU) the atomic throughput benchmark that we described in the previous chapter. This GPU achieves the absolute highest bandwidths among all the GPUs examined (Figure 5.1).

In our experiments, global memory bandwidth does not differ between the PCIe and the NVLink version of the Volta GPU.

Table 5.3: Measured floating-point throughput and bandwidth on the V100 PCIe and NVLink GPUs, using cuBLAS 9.0. We report the highest throughput achieved by matrix multiplication operations encountered while varying the size of matrix from 32 to 8192.

|  |  | V100 PCIe | V100 NVLink |
|---|---|---|---|
| Arithmetic throughput (TFLOPS) | Half precision | 83.03 | 87.42 |
|  | Single precision | 14.03 | 15.53 |
|  | Double precision | 7.07 | 7.83 |
| Bandwidth (GB/s) | L1 cache per SM | 150.56 | 160.85 |
|  | L2 cache per GPU | 2,155 | 2,321 |
|  | shared memory per SM | 151.76 | 168.39 |

## 5.3 Latency

Because Volta's micro-architecture does not vary between its NVLink and PCIe versions, the measured latencies of the two versions are the same on the L1 caches, L2 cache, constant caches, shared memory, and atomic operations.

Global memory access latency appears longer when measured in clock cycles on the NVLink compared to on the PCIe device: 405 vs. 391 clock cycles, respectively. In real terms, however, global memory accesses are marginally faster on the NVLink device than on the PCIe: 26.5 ns vs. 28.3 ns, respectively.

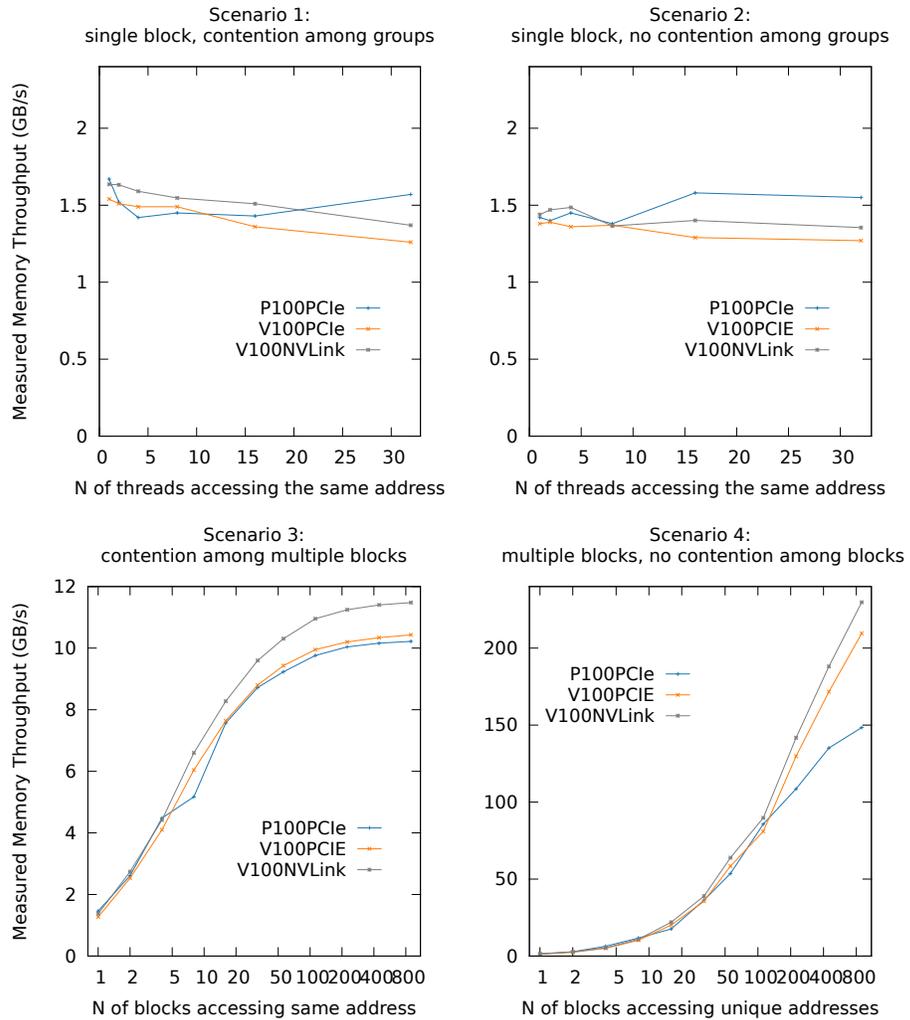

Figure 5.1: Throughput of *atomicAdd()* operations on global memory, measured in four contention scenarios.

# Chapter 6

# Conclusions

Compared to previous architectures, Volta introduces substantial changes in its instruction encoding, memory hierarchy and in the behavior of its processing units.

From Kepler to Volta, the ratio of schedulers to cores grew from 1:48 to 1:16. This change facilitates a higher instruction throughput.

With the newly introduced L0 instruction cache, Volta mitigates the penalty associated with its larger instruction size.

The improved L1 cache offers lower latency and higher bandwidth. Its new replacement policy also reduces cache miss rates when not using shared memory. And the redesigned register banks mitigate bank conflicts.

Thanks to our findings, software designers can optimize their code at the binary level and construct customized assemblers that target Volta, thus obtaining tighter-scheduled code, that delivers higher performance.

Thanks to the memory hierarchy information we disclose, developers can also optimize their code by selecting working sets that match the cache memories at suitable levels, thus reducing their miss rates and improving performance.

# Appendix

In this appendix, we provide the opcodes for common instructions, as encoded in Volta's instruction encoding and, for comparison, in Pascal's encoding as well.

| Name | Pascal | | | | Volta | | | |
|------|------|------|------|------|------|------|------|------|
| | | | | | Floating point instructions | | | |
| FADD | 0101 | 1100 | 0101 | 1 | | 10 | 0010 | 0001 |
| | 0100 | 1100 | 0101 | 1 | | | | |
| | 0011 | 1001 | 0101 | 1 | | | | |
| FCHK | 0101 | 1100 | 1000 | 1 | | 011 | 0000 | 0010 |
| | 0100 | 1100 | 1000 | 1 | | | | |
| | 0011 | 1001 | 1000 | 1 | | | | |
| FCMP | 0101 | 1011 | 1010 | | | | | - |
| | 0101 | 0011 | 1010 | | | | | |
| | 0100 | 1011 | 1010 | | | | | |
| | 0011 | 0111 | 1010 | | | | | |
| FFMA | 0101 | 1001 | 1 | | | 10 | 0010 | 0011 |
| | 0101 | 0001 | 1 | | | | | |
| | 0100 | 1001 | 1 | | | | | |
| | 0011 | 0011 | 1 | | | | | |
| | 0011 | 0010 | 1 | | | | | |
| FMNMX | 0101 | 1100 | 0110 | 0 | | 010 | 0000 | 1001 |
| | 0100 | 1100 | 0110 | 0 | | | | |
| | 0011 | 1001 | 0110 | 0 | | | | |
| | 0011 | 1000 | 0110 | 0 | | | | |
| FMUL | 0101 | 1100 | 0110 | 1 | | 010 | 0010 | 0000 |
| | 0100 | 1100 | 0110 | 1 | | | | |
| | 0011 | 1001 | 0110 | 1 | | | | |
| | 0011 | 1000 | 0110 | 1 | | | | |
| FSET | 0101 | 1000 | | | | 010 | 0000 | 1010 |
| | 0100 | 1000 | | | | | | |
| | 0011 | 0001 | | | | | | |
| FSETP | 0101 | 1011 | 1011 | | | 010 | 0000 | 1011 |
| | 0100 | 1011 | 1011 | | | | | |
| | 0011 | 0111 | 1011 | | | | | |
| | 0011 | 0110 | 1011 | | | | | |
| FSWZADD | 0101 | 0000 | 1111 | 1 | 0 | 1000 | 0010 | 0010 |
| MUFU | 0101 | 0000 | 1000 | 0 | | 011 | 0000 | 1000 |
| RRO | 0101 | 1100 | 1001 | 0 | | | | - |

| Name | Pascal | | | | Volta | | |
|---|---|---|---|---|---|---|---|
| | 0100 | 1100 | 1001 | 0 | | | |
| | 0011 | 1001 | 1001 | 0 | | | |
| | 0011 | 1000 | 1001 | 0 | | | |
| DADD | 0101 | 1100 | 0111 | 0 | 10 | 0010 | 1001 |
| | 0100 | 1100 | 0111 | 0 | | | |
| | 0011 | 1001 | 0111 | 0 | | | |
| | 0011 | 1000 | 0111 | 0 | | | |

Floating point instructions (Cont.)

| Name | Pascal | | | | Volta | | |
|---|---|---|---|---|---|---|---|
| DFMA | 0101 | 1011 | 0111 | | 10 | 0010 | 1011 |
| | 0101 | 0011 | 0111 | | | | |
| | 0100 | 1011 | 0111 | | | | |
| | 0011 | 0111 | 0111 | | | | |
| | 0011 | 0110 | 0111 | | | | |
| DMNMX | 0101 | 1100 | 0101 | 0 | | | - |
| | 0100 | 1100 | 0101 | 0 | | | |
| | 0011 | 1001 | 0101 | 0 | | | |
| | 0011 | 1000 | 0101 | 0 | | | |
| DMUL | 0101 | 1100 | 1000 | 0 | 010 | 0010 | 1000 |
| | 0100 | 1100 | 1000 | 0 | | | |
| | 0011 | 1001 | 1000 | 0 | | | |
| | 0011 | 1000 | 1000 | 0 | | | |
| DSET | 0101 | 1001 | 0 | | | | - |
| | 0100 | 1001 | 0 | | | | |
| | 0011 | 0011 | 0 | | | | |
| | 0011 | 0010 | 0 | | | | |
| DSETP | 0101 | 1011 | 1000 | | 10 | 0010 | 1010 |
| | 0100 | 1011 | 1000 | | | | |
| | 0011 | 0111 | 1000 | | | | |
| | 0011 | 0110 | 1000 | | | | |
| HADD2 | - | | | | 10 | 0011 | 0000 |
| HFMA2 | - | | | | 10 | 0011 | 0001 |
| HMMA2 | - | | | | 0 0010 | 0011 | 0110 |
| HMUL2 | - | | | | 010 | 0011 | 0010 |
| HSETP2 | - | | | | 10 | 0011 | 0100 |
| HSET2 | - | | | | 10 | 0011 | 0011 |
| FSEL | - | | | | 010 | 0000 | 1000 |

Integer Instructions

| Name | Pascal | | | | Volta | | |
|---|---|---|---|---|---|---|---|
| BFE | 0101 | 1100 | 0000 | 0 | | | - |
| | 0100 | 1100 | 0000 | 0 | | | |
| | 0011 | 1001 | 0000 | 0 | | | |
| | 0011 | 1000 | 0000 | 0 | | | |
| BFI | 0101 | 1011 | 1111 | 0 | | | - |
| | 0101 | 0011 | 1111 | 0 | | | |
| | 0100 | 1011 | 1111 | 0 | | | |
| | 0011 | 0111 | 1111 | 0 | | | |
| | 0011 | 0110 | 1111 | 0 | | | |
| FLO | 0101 | 1100 | 0011 | 0 | 011 | 0000 | 0000 |
| | 0100 | 1100 | 0011 | 0 | | | |
| | 0011 | 1001 | 0011 | 0 | | | |
| | 0011 | 1000 | 0011 | 0 | | | |

| Name | Pascal | Volta |
|------|--------|-------|
| IADD | 0101 1100 0001 0 | - |
|  | 0100 1100 0001 0 | |
|  | 0101 1100 0001 0 | |
|  | 0101 1101 0001 0 | |
| IADD3 | 0101 1100 1100 | 010 0001 0000 |
|  | 0100 1100 1100 | |
|  | 0011 1001 1100 | |
|  | 0011 1000 1100 | |
| ICMP | 0101 1011 0100 | - |
|  | 0101 0011 0100 | |
|  | 0100 1011 0100 | |
|  | 0011 0111 0100 | |
|  | 0011 0110 0100 | |
| IMAD | 0101 1010 0 | 10 0010 0100 |
|  | 0101 0010 0 | 10 0010 0101 |
|  | 0100 1010 0 | |
|  | 0011 0100 0 | |

Integer Instructions (Cont.)

| Name | Pascal | Volta |
|------|--------|-------|
| IMADSP | 0101 1010 1 | - |
|  | 0101 0010 1 | |
|  | 0100 1010 1 | |
|  | 0011 0101 1 | |
|  | 0011 0100 1 | |
| IMNMX | 0101 1100 0010 0 | - |
|  | 0100 1100 0010 0 | |
|  | 0011 1001 0010 0 | |
|  | 0011 1000 0010 0 | |
| IMUL | 0011 1000 0011 1 | |
|  | 0100 1100 0011 1 | |
|  | 0011 1001 0011 1 | |
|  | 0011 1000 0011 1 | |
| ISCADD | 0101 1100 0001 1 | |
|  | 0100 1100 0001 1 | |
|  | 0011 1001 0001 1 | |
|  | 0011 1000 0001 1 | |
| ISET | 0101 1011 0101 | - |
|  | 0100 1011 0101 | |
|  | 0011 0111 0101 | |
|  | 0011 0110 0101 | |
| ISETP | 0011 0111 0110 | 010 0000 1100 |
|  | 0100 1011 0110 | |
|  | 0011 0111 0110 | |
|  | 0011 0110 0110 | |
| LEA | 0101 1011 1101 0 | 010 0001 0001 |
|  | 0101 1011 1101 1 | |
|  | 0100 1011 1101 0 | |
|  | 0011 0111 1101 0 | |
|  | 0011 0110 1101 0 | |
|  | 0001 1000 | |
| LOP3 | 0011 11 | 010 0001 0010 |
|  | 0101 1011 1110 0 | |
|  | 0000 001 | |
| LOP | 0101 1100 0100 0 | |
|  | 0100 1100 0100 0 | |

| Name | Pascal | Volta |
|---|---|---|
| | 0011 1001 0100 0 | |
| | 0011 1000 0100 0 | |
| POPC | 0101 1100 0000 1 | 011 0000 1001 |
| | 0100 1100 0000 1 | |
| | 0011 1001 0000 1 | |
| | 0011 1000 0000 1 | |
| SHF | 0101 1011 1111 1 | 10 0001 1001 |
| | 0011 0111 1111 1 | |
| | 0011 1000 1111 1 | |
| | 0011 1001 1111 1 | |
| | 0011 0110 1111 1 | |
| | 0101 1100 1111 1 | |
| SHL | 0101 1100 0100 1 | |
| | 0011 1000 0100 1 | |
| | 0011 1001 0100 1 | |
| | 0100 1100 0100 1 | |
| SHR | 0101 1100 0010 1 | |
| | 0011 1000 0010 1 | |
| | 0011 1001 0010 1 | |
| | 0100 1100 0010 1 | |
| XMAD | 0101 1011 00 | - |
| | 0100 111 | |
| | 0101 0001 0 | |
| | 0011 0111 00 | |
| | 0011 0110 00 | |
| VABSDIFF | - | 10 0001 0100 |
| VABSDIFF4 | - | 10 0001 0101 |
| BREV | - | 011 0000 0001 |

Integer Instructions (Cont.)

| Name | Pascal | Volta |
|---|---|---|
| IABS | - | 010 0001 0011 |
| IDP | - | 010 0010 0110 |
| QSPC | - | 0 0011 1010 1010 |
| BMSK | - | 010 0001 1011 |

Conversion Instructions

| Name | Pascal | Volta |
|---|---|---|
| MOV | 0101 1100 1001 1 | 010 0000 0010 |
| | 0100 1100 1001 1 | |
| | 0011 1001 1001 1 | |
| | 0011 1000 1001 1 | |
| PRMT | 0101 1011 1100 | 10 0001 0110 |
| | 0101 0011 1100 | |
| | 0100 1011 1100 | |
| | 0011 0111 1100 | |
| | 0011 0110 1100 | |
| SEL | 0101 1100 1010 0 | 010 0000 0111 |
| | 0011 1000 1010 0 | |
| | 0011 1001 1010 0 | |
| | 0100 1100 1010 0 | |
| SHFL | 1110 1111 0001 0 | 0 1001 1000 1001 |
| CSET | 0101 0000 1001 1 | - |
| CSETP | 0101 0000 1010 0 | - |

| Name | Pascal | | | | Volta | | |
|---|---|---|---|---|---|---|---|
| PSET | 0101 | 0000 | 1000 | 1 | | | - |
| PSETP | 0101 | 0000 | 1001 | 0 | | | - |
| P2R | 0101 | 1100 | 1110 | 1 | 010 | 0000 | 0011 |
| | 0100 | 1100 | 1110 | 1 | | | |
| | 0011 | 1001 | 1110 | 1 | | | |
| | 0011 | 1000 | 1110 | 1 | | | |
| R2P | 0101 | 1100 | 1111 | 0 | 010 | 0000 | 0100 |
| | 0100 | 1100 | 1111 | 0 | | | |
| | 0011 | 1001 | 1111 | 0 | | | |
| | 0011 | 1000 | 1111 | 0 | | | |
| GETLMEMBASE | - | | | | 0 0011 | 1100 | 0000 |

Load/Store Instructions

| Name | Pascal | | | | Volta | | | |
|---|---|---|---|---|---|---|---|---|
| LD | 100 | | | | 0 | 1001 | 1000 | 0000 |
| LDC | 1110 | 1111 | 1001 | 0 | 0 | 1011 | 1000 | 0010 |
| LDG | 1110 | 1110 | 1101 | 0 | 0 | 0011 | 1000 | 0001 |
| LDL | 1110 | 1111 | 0100 | 0 | 0 | 1001 | 1000 | 0011 |
| LDS | 1110 | 1111 | 0100 | 1 | 0 | 1001 | 1000 | 0100 |
| ST | 101 | | | | 0 | 0011 | 1000 | 0101 |
| STG | 1110 | 1110 | 1101 | 1 | 0 | 0011 | 1000 | 0110 |
| STL | 1110 | 1111 | 0101 | 0 | 0 | 0011 | 1000 | 0111 |
| STS | 1110 | 1111 | 0101 | 1 | 0 | 0011 | 1000 | 1000 |
| ATOM | 1110 | 1101 | | | 0 | 0011 | 1000 | 1010 |
| | 1110 | 1110 | 011 | | 0 | 0011 | 1000 | 1011 |
| | 1110 | 1110 | 1111 | | | | | |
| ATOMS | 1110 | 1100 | | | 0 | 0011 | 1000 | 1100 |
| | 1110 | 1110 | 00 | | 0 | 0011 | 1000 | 1101 |
| | 1110 | 1110 | 010 | | | | | |
| ATOMG | - | | | | 0 | 0011 | 1010 | 1000 |
| | | | | | 0 | 0011 | 1010 | 1001 |
| RED | 1110 | 1011 | 1111 | 1 | 0 | 1001 | 1000 | 1110 |
| CCTL | 1110 | 1111 | 0111 | | 0 | 1001 | 1000 | 1111 |
| MEMBAR | 1110 | 1111 | 1001 | 1 | 0 | 1001 | 1001 | 0010 |
| ERRBAR | - | | | | 0 | 1001 | 1010 | 1011 |

Load/Store Instructions (Cont.)

| Name | Pascal | | | | Volta | | | |
|---|---|---|---|---|---|---|---|---|
| CCTLT | 1110 | 1011 | 1111 | 0 | | | | |
| CCTLL | | | | | 0 | 1001 | 1001 | 0000 |
| MATCH | - | | | | 0 | 0011 | 1010 | 0001 |

Control Instructions

| Name | Pascal | | | Volta | | | |
|---|---|---|---|---|---|---|---|
| BRA | 1110 | 0010 | 0100 | 0 | 1001 | 0100 | 0111 |
| BRX | 1110 | 0010 | 0101 | 0 | 1001 | 0100 | 1001 |
| JMP | 1110 | 0010 | 0001 | 0 | 1001 | 0100 | 1010 |
| JMX | 1110 | 0010 | 0000 | 0 | 1001 | 0100 | 1100 |
| SSY | 1110 | 0010 | 1001 | | | | |
| SYNC | 1111 | 0000 | 1111 1 | | | | - |
| BSYNC | - | | | 0 | 1001 | 0100 | 0001 |

| Name | Pascal | | | Volta | | | |
|---|---|---|---|---|---|---|---|
| WARPSYNC | - | | | | 011 | 0100 | 1000 |
| CAL | 1110 | 0010 | 0110 | | | | - |
| CALL | - | | | | 011 | 0100 | 0011 |
| | | | | 0 | 1001 | 0100 | 0100 |
| JCAL | 1110 | 0010 | 0010 | | | | - |
| PRET | 1110 | 0010 | 0111 | | | | - |
| RET | 1110 | 0011 | 0010 | 0 | 1001 | 0101 | 0000 |
| BRK | 1110 | 0011 | 0100 | | | | - |
| PBK | 1110 | 0010 | 1010 | | | | - |
| CONT | 1110 | 0011 | 0101 | | | | - |
| PCNT | 1110 | 0010 | 1011 | | | | - |
| EXIT | 1110 | 0011 | 0000 | 0 | 1001 | 0100 | 1101 |
| PEXIT | 1110 | 0010 | 0011 | | | | - |
| BPT | 1110 | 0011 | 1010 | | | | |
| BMOV | - | | | 0 | 0011 | 0101 | 0101 |
| | | | | | 011 | 0101 | 0110 |
| | | | | | 011 | 0101 | 0111 |
| YIELD | - | | | 0 | 1001 | 0100 | 0110 |
| RTT | - | | | 0 | 1001 | 0100 | 1111 |
| KILL | - | | | 0 | 1001 | 0101 | 1011 |
| RPCMOV | | | | | 011 | 0101 | 0010 |
| | | | | 0 | 0011 | 0101 | 0011 |
| IDE | - | | | 0 | 1001 | 0101 | 0001 |
| PMTRIG | - | | | 0 | 1000 | 0000 | 0001 |
| BREAK | - | | | 0 | 1001 | 0100 | 0010 |
| BSSY | - | | | 0 | 1001 | 0100 | 0101 |

Other Instructions

| Name | Pascal | | | | | Volta | | | |
|---|---|---|---|---|---|---|---|---|---|
| NOP | 0101 | 0000 | 1011 | 0 | | 0 | 1001 | 0001 | 1000 |
| CS2R | 0101 | 0000 | 1100 | 1 | | 0 | 1000 | 0000 | 0101 |
| S2R | 1111 | 0000 | 1100 | 1 | | 0 | 1001 | 0001 | 1001 |
| B2R | 1111 | 0000 | 1011 | 1 | | 0 | 0011 | 0001 | 1100 |
| BAR | 1110 | 0010 | 0100 | | | | 011 | 0001 | 1101 |
| R2B | 1111 | 0000 | 1100 | 0 | | 0 | 0011 | 0001 | 1110 |
| VOTE | 0101 | 0000 | 1101 | 1 | | 0 | 1000 | 0000 | 0110 |
| | 0101 | 0000 | 1110 | 0 | | | | | |
| TMML | - | | | | | 0 | 1011 | 0110 | 1001 |
| TXD | - | | | | | 0 | 1011 | 0110 | 1100 |
| SGXT | - | | | | | | 010 | 0001 | 1010 |